\documentclass[bibyear]{aa}
\usepackage[varg]{txfonts} 
\usepackage{graphicx}
\usepackage{natbib}
\usepackage{multirow}
\usepackage{amssymb} 
\usepackage{hyperref}
\hypersetup{
colorlinks,
citecolor=blue,
}
\usepackage{xcolor}

\newcommand{\snmad}{$\sigma_{\rm{NMAD}}$}

\defcitealias{Bonoli21}{\bf B21} 

\begin{document}

\bibpunct{(}{)}{;}{a}{}{,} 

\title{The miniJPAS survey: the photometric redshift catalogue}

\author{A.~Hern\'an-Caballero\inst{\ref{CEFCA} \thanks{email: ahernan@cefca.es}}
\and J.~Varela\inst{\ref{CEFCA2}}
\and C.~L\'opez-Sanjuan\inst{\ref{CEFCA2}}
\and D.~Muniesa\inst{\ref{CEFCA}}
\and T.~Civera\inst{\ref{CEFCA}}
\and J.~Chaves-Montero\inst{\ref{DIPC}}
\and L.~A.~D\'iaz-Garc\'ia\inst{\ref{IAA}}
\and J.~Laur\inst{\ref{Tartu}}
\and C.~Hern\'andez-Monteagudo\inst{\ref{ULL},\ref{IAC}}
\and R.~Abramo\inst{\ref{IF/USP}}
\and R.~Angulo\inst{\ref{DIPC}}
\and D.~Crist\'obal-Hornillos\inst{\ref{CEFCA}}
\and R.~M.~Gonz\'alez Delgado\inst{\ref{IAA}}
\and N.~Greisel\inst{\ref{CEFCA}}
\and A.~Orsi\inst{\ref{CEFCA}}
\and C.~Queiroz\inst{\ref{IF/USP},\ref{UFRGS}}
\and D.~Sobral\inst{\ref{Lancaster}}
\and A.~Tamm\inst{\ref{Tartu}}
\and E.~Tempel\inst{\ref{Tartu}}
\and H.~V\'azquez-Rami\'o\inst{\ref{CEFCA2}}
\and J.~Alcaniz\inst{\ref{ON}}
\and N.~Ben\'itez\inst{\ref{IAA}}
\and S.~Bonoli\inst{\ref{CEFCA},\ref{DIPC},\ref{Iker}}
\and S.~Carneiro\inst{\ref{IF/UFB}}
\and J.~Cenarro\inst{\ref{CEFCA2}}
\and R.~Dupke\inst{\ref{ON},\ref{UMich},\ref{UAlab}}
\and A.~Ederoclite\inst{\ref{SaoPaulo}}
\and A.~Mar\'in-Franch\inst{\ref{CEFCA2}}
\and C.~Mendes de Oliveira\inst{\ref{SaoPaulo}}
\and M.~Moles\inst{\ref{CEFCA},\ref{IAA}}
\and L.~Sodr\'e Jr.\inst{\ref{SaoPaulo}}
\and K.~Taylor\inst{\ref{Inst4}}
\and E.~S.~Cypriano\inst{\ref{SaoPaulo}}
\and G.~Mart\'inez-Solaeche\inst{\ref{IAA}}
}

\institute{Centro de Estudios de F\'isica del Cosmos de Arag\'on (CEFCA), Plaza San Juan, 1, E-44001 Teruel, Spain\label{CEFCA}
\and Centro de Estudios de F\'isica del Cosmos de Arag\'on (CEFCA), Unidad Asociada al CSIC, Plaza San Juan, 1, E-44001 Teruel, Spain\label{CEFCA2}
\and Donostia International Physics Centre, Paseo Manuel de Lardizabal 4, E-20018 Donostia-San Sebastian, Spain\label{DIPC}
\and Instituto de Astrof\'{\i}sica de Andaluc\'{\i}a (CSIC), P.O.~Box 3004, E-18080 Granada, Spain\label{IAA}
\and Tartu Observatory, University of Tartu, Observatooriumi 1, 61602 T\~oravere, Estonia\label{Tartu}
\and Departamento de Astrof\'isica, Universidad de La Laguna, E-38206
La Laguna, Tenerife, Spain\label{ULL}
\and Instituto de Astrof\'isica de Canarias, E-38200 La Laguna, Tenerife,
Spain\label{IAC} 
\and Instituto de F\'isica, Universidade de S\~ao Paulo, Rua do Mat\~ao 1371, CEP 05508-090, S\~ao Paulo, Brazil\label{IF/USP}
\and Departamento de Astronomia, Instituto de F\'isica, Universidade Federal do Rio Grande do Sul (UFRGS), Av. Bento Gon\c{c}alves 9500,
Porto Alegre, R.S, Brazil\label{UFRGS}
\and Department of Physics, Lancaster University, Lancaster, LA1 4YB, UK\label{Lancaster}
\and Observat\'orio Nacional, Minist\'erio da Ci\^encia, Tecnologia, Inova\c{c}\~ao e Comunica\c{c}\~oes, Rua General Jos\'e Cristino, 77, S\~ao Crist\'ov\~ao, 20921-400, Rio de Janeiro, Brazil\label{ON}
\and Ikerbasque, Basque Foundation for Science, E-48013 Bilbao, Spain\label{Iker}
\and Instituto de F\'isica, Universidade Federal da Bahia, 40210-340, Salvador, BA, Brazil\label{IF/UFB}
\and Department of Astronomy, University of Michigan, 311 West Hall, 1085 South University Ave., Ann Arbor, USA\label{UMich}
\and Department of Physics and Astronomy, University of Alabama, Box 870324, Tuscaloosa, AL, USA\label{UAlab}
\and Departamento de Astronomia, Instituto de Astronomia, Geof\'isica e Ci\^encias Atmosf\'ericas da USP, Cidade Universit\'aria, 05508-900, S\~ao Paulo, SP, Brazil\label{SaoPaulo}
\and Instruments4\label{Inst4}
}

\date{Accepted ........ Received ........;}

\abstract {
MiniJPAS is a $\sim$1 deg$^2$ imaging survey of the AEGIS field in 60 bands, performed to demonstrate the scientific potential of the upcoming Javalambre-Physics of the Accelerating Universe Astrophysical Survey (J-PAS).
Full coverage of the 3800--9100 \AA{} range with 54 narrow-band filters, in combination with 6 optical broad-band filters, allow for extremely accurate photometric redshifts (photo-$z$), which applied over areas of thousands of square degrees will enable new applications of the photo-$z$ technique such as measurement of baryonic acoustic oscillations.
In this paper we describe the method used to obtain the photo-$z$ included in the publicly available miniJPAS catalogue, and characterise the photo-$z$ performance. 
We build 100 \AA{} resolution photo-spectra from the PSF-corrected forced-aperture photometry. Systematic offsets in the photometry are corrected by applying magnitude shifts obtained through iterative fitting with stellar population synthesis models.
We compute photo-$z$ with a customised version of {\sc LePhare}, using a set of templates optimised for the J-PAS filter-set. 
We analyse the accuracy of miniJPAS photo-$z$ and their dependence on multiple quantities using a subsample of 5,266 galaxies with spectroscopic redshifts from SDSS and DEEP, that we find to be representative of the whole $r$$<$23 miniJPAS sample.
Formal 1-$\sigma$ uncertainties for the photo-$z$ that are calculated with the $\Delta\chi^2$ method underestimate the actual redshift errors. The $odds$ parameters has the stronger correlation with $\vert\Delta z\vert$, and  accurately reproduces the probability of a redshift outlier ($\vert\Delta z\vert$$>$0.03) irrespective of the magnitude, redshift, or spectral type of the sources. 
We show that the two main summary statistics characterising the photo-$z$ accuracy for a population of galaxies (\snmad{} and $\eta$) can be predicted by the distribution of $odds$ in such population, and use this to estimate them for the whole miniJPAS sample.
At $r$$<$23 there are $\sim$17,500 galaxies per deg$^2$ with valid photo-$z$ estimates, of which $\sim$4,200 are expected to have $\vert\Delta z\vert$$<$0.003. The typical error is \snmad=0.013 with an outlier rate $\eta$=0.39. The target photo-$z$ accuracy \snmad=0.003 is achieved for $odds$$>$0.82 with $\eta$=0.05, at the cost of decreasing the density of selected galaxies to $n$$\sim$5,200 deg$^{-2}$ (of which $\sim$2,600 have $\vert\Delta z\vert$$<$0.003).
}


\titlerunning{The photo-$z$ catalogue of miniJPAS}
\authorrunning{A. Hern\'an-Caballero et al.}
\maketitle

\section{Introduction} 

The idea of using multi-wavelength photometry to estimate the redshift of galaxies (photometric redshifts, photo-$z$ in the following) was first proposed as a last resort technique to obtain redshifts for sources deemed too faint for spectroscopy to be feasible or economical \citep{Baum62,Couch83}. Photo-$z$ became increasingly useful with the advent of deep multi-wavelength imaging surveys, starting with the Hubble Deep Field \citep[e.g.][]{Lanzetta96,Mobasher96,Fernandez-Soto99}.
The main advantage of photo-$z$ over spectroscopic redshifts (spec-$z$) is the ability to obtain redshifts for all the sources detected in an imaging survey without pre-selection. Combined with large CCD detectors, this implies an increase in the survey speed of orders of magnitude compared to the most advanced multi-object spectrographs \citep{Blake05}.
 
The main drawback of the photo-$z$ method is the accuracy of its redshift estimates, typically much lower compared to spec-$z$, imposing hard constraints on their range of application.
However, in the last few years photo-$z$ techniques have matured enough to promote a shift in the redshift strategy of many current and upcoming surveys, which now have photo-$z$ as their primary method for distance determination. Spectroscopy is still essential for obtaining calibration and validation samples needed to fine-tune the photo-$z$ machinery, or to follow up on particularly interesting sources. However, relying on photo-$z$ for an overwhelming majority of targets allows for complete, flux-limited samples containing many millions of galaxies, which opens up entirely new applications such as baryonic acoustic oscillation (BAO) measurements \citep{Blake05,Angulo08,Chaves-Montero18}.

Essential to this rise to preeminence of photo-$z$ has been their huge improvement in accuracy and reliability over time, in particular on imaging surveys designed with photo-$z$ in mind, which split the optical range into increasingly large numbers of ever narrower bandpasses. The first wide-area survey of the photo-$z$ era was the Sloan Digital Sky Survey \citep[SDSS;][]{York00}, with only 5 (carefully designed) broadband filters. COMBO-17 \citep{Wolf03} and Subaru COSMOS 20 \citep{Taniguchi07,Taniguchi15} raised the number of filters to 17 and 20, respectively, including broad- and medium- or narrow-band filters. The Advanced, Large, Homogeneous Area, Medium- Band Redshift Astronomical1 (ALHAMBRA) survey \citep{Moles08} observed $\sim$4 deg$^2$ in 20 optical medium-band filters (FWHM$\sim$300 \AA) combined with JHKs near-infrared imaging. 
The Survey for High-z Absorption Red and Dead Sources \citep[SHARDS;][]{Perez-Gonzalez13} imaged the Hubble Deep Field in 25 contiguous optical filters with FWHM$\sim$170 \AA{}, and the Physics of the Accelerating Universe Survey (PAUS) observed the COSMOS field in 40 filters with FWHM$\sim$130 \AA{} \citep{Eriksen19}. 

A new landmark of the application of photo-$z$ will be reached with the Javalambre-Physics of the Accelerating Universe Astrophysical Survey \citep[J-PAS;][]{Benitez09,Benitez14}, specifically designed to achieve the high redshift accuracy required to perform BAO measurements over a wide redshift range.
Using 54 narrow-band filters (FWHM$\sim$145 \AA) complemented with two broadband filters at both extremes of the optical range, J-PAS imaging will effectively obtain a low-resolution spectrum (photo-spectrum) for every 0.46\arcsec pixel in the sky over thousands of square degrees. 
The J-PAS survey will be performed by the 2.5-m Javalambre Survey Telescope (JST), equipped with a 1.2 Gigapixel camera (JPCam) with a field of view of 4.2 deg$^2$. 

To test the performance of JST and begin scientific operation prior to the installation of JPCam, a pathfinder camera (PF) with a single 9k x 9k CCD was installed. To prove the scientific potential of J-PAS, a small survey of $\sim$1 deg$^2$ (miniJPAS) was carried out with PF on the AEGIS field using the 56 J-PAS filters as well as 4 broadband filters ($u$, $g$, $r$, $i$).
MiniJPAS observations span 4 overlapping pointings along the Extended Groth Strip (EGS), reaching the depth planned for J-PAS (5-$\sigma$ limits between $\sim$21.5 and 22.5 for the narrow-band filters and $\sim$24 for the broad-band filters in a 3\arcsec aperture). A detailed description of the miniJPAS observations, data reduction, and calibration is presented in \citet[][hereafter \citetalias{Bonoli21}]{Bonoli21}. Fully reduced images and source catalogues are publicly available at the CEFCA catalogues portal.\footnote{\url{http://archive.cefca.es/catalogues/minijpas-pdr201912}} 

In this paper, we present the photometric redshift catalogue for miniJPAS. Section 2 outlines the miniJPAS observations and data reduction, and describes the photometric measurements used for photo-$z$ calculation and the spectroscopic redshifts. Section 3 discusses our photometric recalibration procedure. Section 4 offers an overview of the photo-$z$ code, the templates used, and its outputs, while Sect. 5 discusses the sources of error in photo-$z$ estimates. In Sect. 6, we present the main results on the redshift distribution and photo-$z$ accuracy for miniJPAS galaxies and its dependence on multiple quantities. Finally, Sect. 7 summarises our conclusions. All magnitudes are presented in the AB system.

\section{Data}\label{sec:data}

\subsection{observations and data reduction}

The observations and data reduction of miniJPAS are described in detail in Sect. 2 and Sect. 3 of \citetalias{Bonoli21}. Here we provide a brief summary. 
The miniJPAS covers the EGS in 4 overlapping pointings, for a total area of $\sim$1 deg$^2$. Each pointing was observed with a minimum of 4 exposures per filter, with a dithering of 10\arcsec along both CCD axes.
The readout of the CCD was done in 2$\times$2 binning mode for the narrow-band filters (0.46\arcsec~pix$^{-1}$) and 1$\times$1 for broad-band (0.23\arcsec~pix$^{-1}$). 
Individual exposure times were 120 s for the 56 narrow-band filters as well as the $u$ filter, but only 30 s for the other broadband filters to prevent saturation. Total exposure times range from 480 s to 3240 s, depending on the filter and pointing (see Table A.1 in \citetalias{Bonoli21} for details). 

Data reduction for the individual images includes the standard bias and over-scan subtraction, trimming, flat fielding, and illumination correction. Also, some issues specific to the Pathfinder camera (which unlike JPCam, was not designed specifically for the JST/T250) required additional corrections for vignetting, background patterns, and fringing (see \citetalias{Bonoli21}).

Astrometric calibration was performed with {\sc Scamp} \citep{Bertin06} using the Gaia DR2 catalogue as reference. The astrometric solution for the individual images has a rms of $\sim$0.035\arcsec with respect to Gaia. Coaddition of the individual images was performed with {\sc Swarp} \citep{Bertin02}, with all the images resampled to the fiducial pixel scale of the camera (0.23\arcsec). 

The average 5-$\sigma$ depth of the coadded narrow-band images ranges from $\sim$21.5-22.0 AB magnitudes in the reddest filters to $\sim$22.5-23.5 in the bluest ones. For broad-band filters, it is $u$$\sim$22.8, $g$$\sim$24.0, $r$$\sim$23.8, and $i$$\sim$23.2.

The PSF FWHM of the coadded images ranges from $\sim$0.6\arcsec to $\sim$2.0\arcsec, with most of them below the 1.5\arcsec mark. The images for the reference band ($r$) have the lowest PSF FWHM, averaging 0.7\arcsec in the four pointings.

\subsection{Photometry}\label{subsec:photometry}

Source detection and extraction on the reduced miniJPAS images was performed with {\sc SExtractor} \citep{Bertin96}. Aperture photometry was obtained in both dual-mode and single-mode for several types and aperture sizes and calibrated using an adaptation of the method presented in \citet[][see Sects.~3.3 and 3.4 in \citetalias{Bonoli21} for details]{Lopez-Sanjuan19b}. 

In dual-mode, the extraction aperture is defined in a reference band (in our case the detection band $r$) and used to perform forced photometry on the images in all the other filters).
{\sc SExtractor} computes total magnitudes using the auto (MAG\_AUTO), isophotal (MAG\_ISO), and Petrosian (MAG\_PETRO) apertures. These apertures are defined individually for each source based on the $r$ band so that the fraction of the (estimated) total flux of the galaxy enclosed is constant, irrespective of the galaxy size and surface brightness profile. A constant scaling factor then converts the fluxes integrated in the apertures to total fluxes.

However, accurate total fluxes are much less important than accurate colour indices for the purpose of photo-$z$ estimation. Obtaining accurate colours requires to maximise the signal-to-noise ratio (S/N) within the extraction aperture and to compensate for PSF variation among the images in different filters. In order to achieve this, we use 
PSF corrected magnitudes (MAG\_PSFCOR) obtained following the method presented in \citet{Molino19}, which is based on {\sc ColorPro} \citep{Coe06}.

Very briefly, PSFCOR magnitudes are obtained by extracting the flux in
a small aperture, defined as the Kron aperture \citep{Kron80} with semi-major axis equal to 1 Kron radius in the reference band. This is half the size of a standard AUTO aperture in {\sc SExtractor} (2 Kron radii). The resulting magnitude is known as the ``restricted AUTO'' or just ``restricted'' magnitude \citep{Molino17,Molino19}. 
For the reference band, the PSFCOR magnitude is simply the restricted magnitude.
For bands with wider PSF than the reference band, a correction term for PSF broadening is applied, as follows:

\begin{equation}
\textrm{PSFCOR}_j = \textrm{REST}_j + \textrm{REST}_r - \textrm{REST($j$)}_r
\end{equation}

\noindent where \textrm{REST}$_j$ and \textrm{REST}$_r$ are the magnitudes measured on the restricted aperture for the images in band $j$ and the reference band, respectively, while \textrm{REST($j$)}$_r$ is measured on the reference image after convolution to the same PSF of the image in band $j$.

If, on the contrary, band $j$ has a narrower PSF compared to the reference band, then the PSFCOR magnitude is simply the restricted magnitude measured on the image after convolution to the same PSF of the reference band.
We emphasise that PSFCOR magnitudes are not total magnitudes. They underestimate the total flux of galaxies by $\sim$0.5 mag, on average, with respect to AUTO magnitudes \citep[see also][]{GonzalezDelgado21}.

Hereafter, we will omit the explicit reference to the aperture type when discussing magnitudes, colours, or magnitude offsets. Unless otherwise stated, it will be implicit that AUTO magnitudes are used whenever the flux or luminosity in a single band is needed (e.g.: selection of a flux-limited sample, redshift priors) while PSFCOR magnitudes are used for SED fitting and colour-dependent quantities.

The photometry is corrected for atmospheric extinction as part of the calibration process \citep[see ][for details]{Lopez-Sanjuan19b}. We also correct for Galactic extinction using the Milky Way dust maps from Bayestar17 \citep{Green18} with extinction coefficients $k_\lambda$ computed using the prescription in \citet{Whitten19} for the extinction law of \citet{Schlafly16}.

\subsection{Flags}

The miniJPAS catalogue includes the column FLAGS, which contains 
information on whether each source is affected, in each of the bands, by a number of issues that may impair or invalidate the photometry. 
The FLAGS value is an integer that encodes several binary flags (see \citetalias{Bonoli21} for the details), including the {\sc SExtractor} flags (which alert for close neighbours, blending, saturation, truncation, etc) as well as two additional flags that mark sources duplicated in another tile or that are known to be variable. 
Sources that have no apparent issues in their photometry in a particular band have FLAGS=0 for that band. Sources with FLAGS$>$0 in at least one band represent $\sim$30\% of the miniJPAS catalogue. 

We compute photo-$z$ for every source brighter than $r$$<$24 and with FLAGS$<$4 in the detection band (this removes saturated and truncated sources as well as those with incomplete or corrupted data in the extraction aperture).
The FLAGS values are used to choose which bands are considered for the photo-$z$ calculation.
In the general case, we select only those bands with FLAGS=0 (meaning none of the {\sc SExtractor} flags or additional flags is raised).
However, if a source has FLAGS$>$0 in more than 50\% of the bands, we relax this condition by requiring FLAGS$<$4, which dismisses the values indicating close neighbours or blending.

The impact of the photometric issues signalled by the flags on the photo-$z$ accuracy is hard to predict as it depends on many factors, including the specific bands affected, the error introduced in the photometric measurement, the SED of the source, and its redshift. In general, the photo-$z$ accuracy is degraded in sources with flags, and, in particular, the rate of catastrophic errors (outliers) is substantially higher. 

\begin{figure} 
\begin{center}
\includegraphics[width=8.4cm]{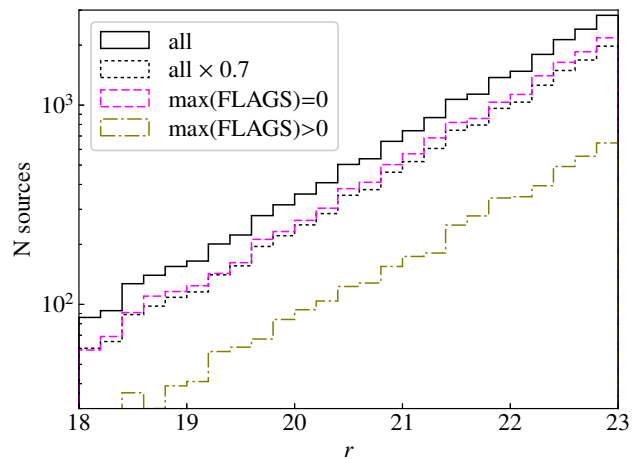}
\end{center}
\caption[]{Distribution of $r$-band magnitudes for the whole miniJPAS sample (black solid line), flagged sources (olive dot-dashed line), and non-flagged sources (magenta dashed line). The distribution for the whole sample scaled by a factor 0.7 is shown for reference (black dotted line).\label{fig:Nsources-mag-flags}}
\end{figure} 

Except for saturation, all the conditions signaled by the photometry flags depend mainly on the location of the source within the image, not its intrinsic properties. As a consequence, which sources are flagged should not depend on their brightness, apparent size, redshift, or spectral type. Figure \ref{fig:Nsources-mag-flags} confirms this to actually be the case for brightness, with $\sim$30\% of sources being flagged irrespective of their magnitude.

Because of this, we consider the $\sim$70\% of sources with no flags to be representative of the full miniJPAS sample. We will use this sub-sample to discuss the properties of the population of miniJPAS sources throughout the paper.

\subsection{Star/galaxy classification\label{sec:star-galaxy-class}}

The miniJPAS catalogue includes the results of different methods for separating stars from galaxies. The most basic approach is the CLASS\_STAR parameter from {\sc SExtractor}, which compares the spatial profile of the source with the expectation for a point source.
A more realistic morphological classification is given by MORPH\_PROB\_STAR which gives the probability of a source being a star given its spatial profile and a prior probability based on its $r$-band magnitude \citep[see][for details]{Lopez-Sanjuan19a}.
TOTAL\_PROB\_STAR combines MORPH\_PROB\_STAR with parallax information from Gaia for a more robust determination.
Finally, ERT\_PROB\_STAR provides classification using the Extremely Randomised Trees machine learning method, which uses morphological and photometric parameters \citep{Baqui20}.

Although ERT\_PROB\_STAR has the highest success rate in separating stars from galaxies \citep{Bonoli21,Baqui20}, we choose TOTAL\_PROB\_STAR as the preferred measurement of the probability of being a star, $P_S$, through the paper. This is because TOTAL\_PROB\_STAR has the advantage of not relying on the SED of the galaxy, which is important for obtaining the redshift distribution of galaxies in a consistent way, as we show below. 

\begin{figure} 
\begin{center}
\includegraphics[width=8.4cm]{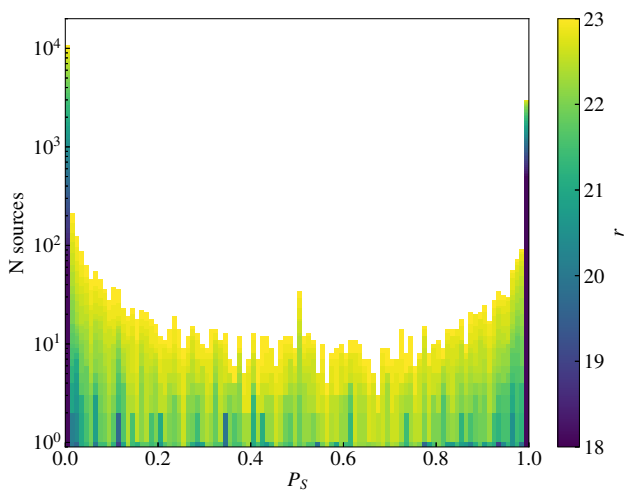}
\end{center}
\caption[]{Distribution of the probability of being a star, $P_S$, for miniJPAS sources as a function of the (colour-coded) limiting magnitude of the sample. The spike at $P_S$=0.5 is due to sources with insufficient S/N for classification in the detection band, which get $P_S$=0.5 by default.\label{fig:pstar-distrib}}
\end{figure} 

The distribution of $P_S$ is strongly bimodal, with most sources concentrated near the extremes of its range, at $P_S$$\sim$0 (very likely to be a galaxy) or $P_S$$\sim$1 (star, see Fig. \ref{fig:pstar-distrib}). In particular, 96.8\% of the sources brighter than $r$=22 have either $P_S$$<$0.01 or $P_S$$>$0.99.
However, at faint magnitudes the distinction becomes more uncertain, with many sources taking intermediate values of $P_S$.
As a consequence, applying a cut in $P_S$ to tell apart galaxies from stars results in contamination and incompleteness in both samples. Any determination of statistical properties of miniJPAS galaxies based on such partitioning, (including the redshift distribution) would be biased at faint magnitudes.

To solve this issue, we compute photo-$z$ for all miniJPAS sources regardless of the morphological classification. For each source we obtain $P$($z$|G), the redshift probability distribution conditional to the source being a galaxy, which depends on its SED alone. If a source is known to be a star or quasar ($P_S$ = 1) then the value of $P$($z$|G) is meaningless. However, by computing $P$($z$|G) independently of any morphological information and applying the morphological classification as a posterior, we can easily estimate redshift-dependent statistics for the population of miniJPAS galaxies that account for the uncertainty in the classification of individual sources (see Sect. \ref{sec:results}). 

\subsection{Spectroscopic redshifts}

\begin{figure} 
\begin{center}
\includegraphics[width=8.4cm]{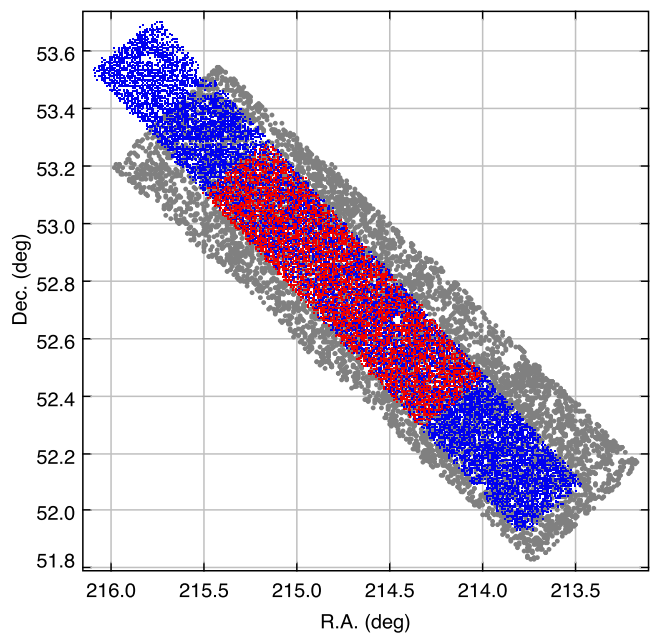}
\end{center}
\caption[]{Footprint of the miniJPAS and DEEP2/DEEP3 spectroscopic observations. Grey circles represent miniJPAS sources with $r$$<$23, while blue and red dots represent sources targeted for spectroscopy by the DEEP2 and DEEP3 surveys, respectively.\label{fig:zspec-footprint}}
\end{figure} 

The Deep Extragalactic Evolutionary Probe 2 (DEEP2) and 3 (DEEP3) galaxy redshift surveys \citep{Davis03,Cooper11,Newman13} cover about half of the area of the miniJPAS survey (Fig. \ref{fig:zspec-footprint}).
Both surveys were performed with the DEIMOS multi-object spectrograph on the Keck II telescope. DEEP2 spectra were obtained with the 1200 lines/mm grating, covering the $\sim$6500--9100 \AA{} range with a spectral resolution of $R$$\sim$5000,
while DEEP3 used the 600 lines/mm grating, allowing for a wider spectral coverage ($\sim$4550--9900 \AA) at $R$$\sim$2500. 

We retrieved the combined DEEP2/3 redshift catalogue for the EGS\footnote{http://deep.ps.uci.edu/deep3/zcat\_archive/\\
alldeep.egs.uniq.2012jun13.fits.gz}, which contains a total of 23822 unique sources.
Targets for DEEP2 observations were selected at random from the $R$$<$24.1 flux limited catalogue of \citet{Coil04}, which covers the entirety of the EGS. By contrast, DEEP3 covers only the central part of the EGS. While most DEEP3 targets are also selected from the $R$$<$24.1 sample, additional sources were targeted based on detection at other wavelengths, such as far-infrared or X-rays, including some $R$$>$24.1 sources. We remove these sources from the catalogue to keep the spectroscopic sample representative of the population of  $R$$<$24.1 sources. We also require that sources have secure redshifts (ZQUALITY $>$=3) and spectral classification as galaxy. There are 15,222 sources meeting all these criteria. 

We perform a match by coordinates between this sample and the miniJPAS catalogue using a search radius of 1.5''. We find spectroscopic counterparts for 4,825 out of 20,962 miniJPAS sources brighter than $r$=23, at an average matching distance of 0.12''.

We also match the miniJPAS catalogue with the spectroscopic redshift catalogue from SDSS DR12, which covers the entirety of the miniJPAS footprint. Again, we keep only matches for sources with secure spectroscopic redshifts (zwarning = 0) and spectroscopic classification as galaxy. 564 sources meet these criteria, including 123 that also have redshifts from DEEP. In all the galaxies in common between SDSS and DEEP their redshift measurements agree to within $\vert\delta z\vert$$<$0.001, the median $\vert\delta z\vert$ being 0.00012.

\begin{figure} 
\begin{center}
\includegraphics[width=8.4cm]{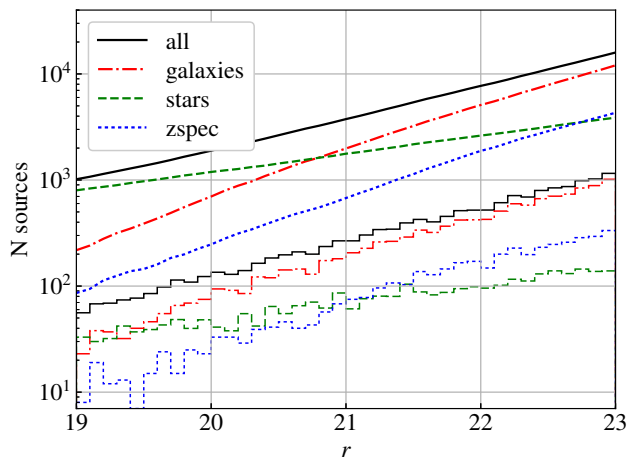}
\end{center}
\caption[]{Distribution of miniJPAS sources as a function of $r$-band magnitude for all sources (black solid lines), stars (green dashed lines), galaxies (red dot-dashed lines), and galaxies with spectroscopic redshifts (blue dotted lines). Smooth lines show cumulative counts while histograms show counts in bins of 0.1 magnitudes.\label{fig:Nmag}}
\end{figure}

Figure \ref{fig:Nmag} shows the distribution of sources of all types, galaxies, and galaxies with spectroscopic redshifts as a function of the $r$-band magnitude in the miniJPAS catalogue. The distribution for all sources is computed without taking into account the morphological classification and is dominated by stars at $r$$<$19.5. The distributions for stars and galaxies are obtained by weighting each source with its $P_S$ and $P_G$, respectively. Finally, the distribution for galaxies with spectroscopic redshifts is computed from the subsample with matches in our spectroscopic catalogue. 

The fraction of miniJPAS galaxies with spectroscopic redshifts is roughly constant at $\sim$25\% within the range 19$<r<$23. This fraction increases to $\sim$50\% if we consider only sources within the DEEP footprint (the number of SDSS spectra for $r$$>$19 galaxies is negligible).

\section{Recalibration}\label{subsec:recalibration}

\begin{figure} 
\begin{center}
\includegraphics[width=8.4cm]{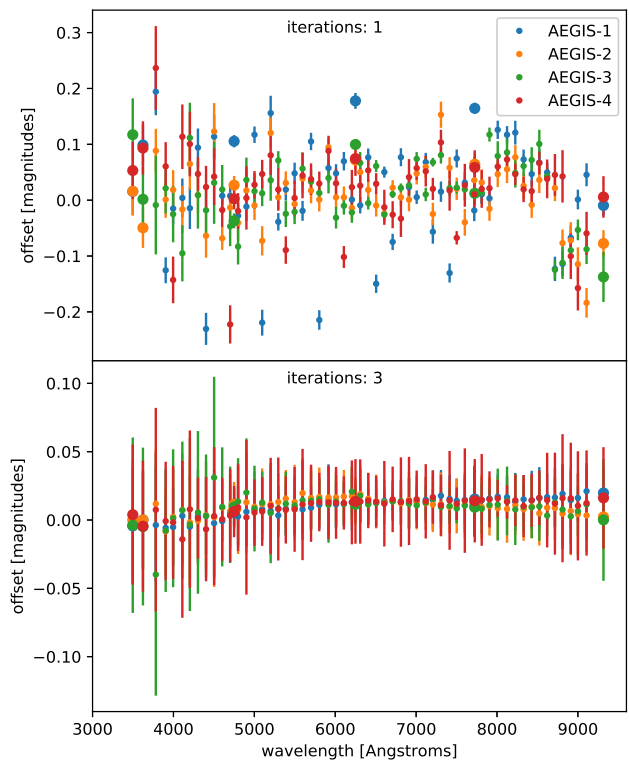}
\end{center}
\caption[]{Magnitude residuals between the observed and synthetic photometry after the first (top) and third (bottom) iterations of the recalibration procedure. Symbols indicate the median residual while error bars enclose the 16$^{th}$ to 84$^{th}$ percentile ranges. Small symbols correspond to photometry the narrow-band J-PAS filters while large ones represent broadband filters (uJAVA, uSDSS, gSDSS, rSDSS, iSDSS) and the long-pass J1007 filter.\label{fig:recalib}}
\end{figure} 

The observed colours of galaxies may be affected by systematics in the photometry originating from errors in the absolute flux calibration of the images, in the correction for Galactic extinction, or aperture effects introduced by PSF variation among images in different bands.
Because accurate galaxy colours are essential for photometric redshifts, 
many SED-fitting photo-$z$ codes include a pre-processing stage dedicated to 
computing magnitude offsets for each band that minimise the average colour differences between the observed photometry and synthetic photometry extracted from the spectral templates \citep[e.g.][]{Ilbert06,Molino14,Eriksen19}. We call this process zero-point recalibration, or recalibration for short. 
Since the spectral templates are also affected by calibration uncertainties, this procedure has the advantage of putting the observed photometry and the templates in the same ``photometric system''.

A major shortcoming of this approach is that the template set usually contains only a few ``archetype'' templates representative of the broad spectral types expected to be found in the sample. For datasets such as miniJPAS, with observations in many bands, this implies that poor fits (high $\chi^2$ values) are obtained for many sources (in particular the ones with highest S/N) even after the recalibration offsets have converged, because the template set is not big and diverse enough to reproduce all the variety of spectral features found in individual galaxies. Discrepancies between the best-fitting template and the actual spectrum of the galaxy increase the dispersion in the magnitude offsets calculated from different galaxies for each band, implying a more uncertain correction.

To overcome this issue, we instruct the photo-$z$ code to skip the recalibration step\footnote{in {\sc LePhare} this is done by setting AUTO\_ADAPT=NO in the configuration file} and instead we perform the recalibration separately, using a custom routine. 
This allows us to use a very large grid of spectral templates, built from stellar population synthesis models, which can reproduce the SED of miniJPAS galaxies to much higher accuracy. 

\subsection{Grid of models}

We generate a grid of 90,720 synthetic spectra with the python implementation of {\sc cigale}, that is described in \citet{Boquien19}. Each one represents the theoretical spectrum of a galaxy composed of two stellar populations (namely 'young' and 'old'). The old population is assumed to have a delayed-exponential star formation history (SFH), with age between 2 and 10 Gyr and e-folding timescale between 0.5 and 2 Gyr. The SFH for the young population is also a delayed-exponential, with age between 0.1 and 1 Gyr and e-folding timescale between 50 and 200 Myr. The fractional contribution of the young population to the total stellar mass of the galaxy may take values from 0\% to 10\%.

The synthetic spectrum for each population is generated by integration over time of a library of simple stellar populations (SSPs). We choose the high spectral resolution version of the library of \citet{Bruzual03}, with a \citet{Chabrier03} initial mass function (IMF), and stellar metallicities $Z$/$Z_\odot$ = 0.2, 0.4, and 1.0.
The nebular emission is modelled with the grid of nebular templates from \citet{Inoue11}, which were generated using CLOUDY 13.01 \citep{Ferland98,Ferland13}. The metallicity of the gas is assumed to be the same as that of the stellar component. To keep the total number of models manageable, we restrict values of the ionisation parameter to the range $\log$ U = [-3,-1] in steps of 0.5. The fraction of Lyman continuum photons reprocessed into nebular emission (that is, not absorbed by dust or escaping into the intergalactic medium) ranges from 5\% to 100\%.
The attenuation law is \citet{Calzetti00}, and the colour excess E(B-V) of both stellar components is fixed at 0.44 times that of the nebular component (which takes values from 0 to 0.5 mag in steps of 0.1).

\subsection{Computation of offsets}

We perform the recalibration using miniJPAS galaxies with secure spectroscopic redshifts and spectroscopic classification as ``galaxy'' from either SDSS or DEEP. Since complete and accurate SEDs are essential for the recalibration procedure, we exclude sources fainter than $r$=22 or with flags in the photometry of any of the 60 bands. 

We obtain recalibration offsets independently for each of the four pointings of the miniJPAS survey using $\sim$500 galaxies per pointing.
For each galaxy, the model rest-frame spectra in the grid are redshifted to the observed frame of the galaxy. Then synthetic photometry is obtained by convolving the redshifted spectra with the transmission curves of the filters,
and a scaling factor is found that provides the best fit (minimum $\chi^2$) between the observed and synthetic SEDs. 
We select as the best model for each galaxy the one that produces the absolute minimum $\chi^2$ among the 90,720 models in the grid.
Then, for every pointing, we compute the systematic offset in the photometry for band $j$ as:

\begin{equation}
\delta m(j) = \textrm{median} \big{\{} m_i^{obs}(j) - m_i^{synth}(j) \big{\}}
\end{equation}

\noindent where $m_i^{obs}$($j$) and $m_i^{synth}$($j$) are, respectively, the observed and synthetic magnitudes in band $j$ for the $i$-th galaxy.
The observed photometry is adjusted by subtracting the offsets just computed:

\begin{equation}
m_i^{corr}(j) = m_i^{obs}(j) - \delta m(j)
\end{equation}

\noindent We repeat the SED-fitting with the grid of models for the updated photometry and recompute the offsets iteratively, until all new additional offsets are smaller than 0.001 magnitudes, which in practice happens after 3 or 4 iterations (Fig. \ref{fig:recalib}). The cumulative offsets resulting from the addition of the successive offsets in all the iterations, $\Delta m$($j$) = $\delta m$($j$) + $\delta m'$($j$) + \ldots + $\delta m^n$($j$), are the final recalibration corrections, which we apply to the original photometry for all the sources in the pointing. 

We estimate the uncertainty in the recalibration offsets as the uncertainty in the median of the residuals after the last iteration, assuming a normal distribution:

\begin{equation}
\sigma\big{(}\Delta m(j)\big{)} = \sqrt{\frac{\pi}{2N}} \sigma \big{\{} m_i^{obs}(j) - \Delta m(j) - m_i^{synth}(j) \big{\}}
\end{equation}

\noindent where $N$ is the number of galaxies used and $m_i^{obs}$($j$) - $\Delta m$($j$) is the magnitude in band $j$ of the $i$-th galaxy after recalibration.

The recalibration offset computed for a given band differs significantly from one pointing to another, mostly due to differences in the PSF FWHM of the images. Table \ref{table:zp_offsets} indicates the offsets and their uncertainties for PSFCOR magnitudes in the 60 bands for the the four miniJPAS pointings. 

\subsection{Validation of recalibration offsets}

The recalibration procedure may raise some legitimate concerns about its robustness. One possible source of problems is degeneracy between offsets caused by systematic errors in the photometry and those in the spectral templates. If most of the galaxies in the recalibration sample are within a small redshift range, any given spectral feature will appear most of the time in only a few bands. This has the potential of biasing the recalibration if the best-fitting models for these galaxies systematically under- or over-predict the intensity of the feature. Emission lines, whose intensity is particularly hard to predict from the observed continuum, are one clear example. 
 
In order to check whether this degeneracy is an issue for the miniJPAS sample,
we split the recalibration sample for each pointing into two subsamples, one containing all the galaxies with redshift below the median of the sample and the other with the ones above. We perform the recalibration separately for the two subsamples and compare the resulting offsets. We find that the offsets calculated for the two subsamples are consistent within their uncertainties.

Another sensible concern is the unicity of the results from the recalibration procedure. 
To test whether the recalibration converges to the same corrected photometry irrespective of the systematic offsets, we modify the observed photometry by applying random shifts between -0.2 and +0.2 magnitudes (selected from an uniform distribution). The same shift $x$($j$) is applied to all galaxies for each band $j$ to simulate a systematic offset.

We compute recalibration corrections, $\Delta m^*$($j$), for this modified photometry using the exact same method as before. If the recalibration can compensate for these additional systematic shifts, then we expect the difference between the original and the new recalibration offsets to match the shift applied to the photometry:

\begin{equation}
\Delta m(j) - \Delta m^*(j) \approx x(j)
\end{equation}

\begin{figure} 
\begin{center}
\includegraphics[width=8.4cm]{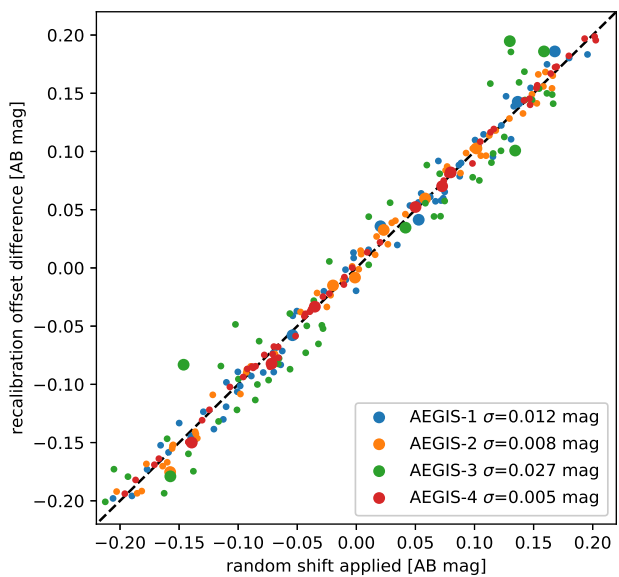}
\end{center}
\caption[]{Comparison between the arbitrary shifts introduced in the input photometry for individual bands (X axis) and the resulting change in the recalibration offsets obtained (Y axis). Symbols as in Fig. \ref{fig:recalib}. The dashed line marks the 1:1 relation expected when the recalibration process compensates perfectly systematic offsets in the photometry.\label{fig:validate-recalibration}}
\end{figure} 

Figure \ref{fig:validate-recalibration} compares the random shifts introduced, $x$($j$), with the resulting change in the recalibration offsets, $\Delta m$($j$)-$\Delta m^*$($j$), for all the bands and pointings. The dispersion around the 1:1 relation is $\sim$0.01 mag (except for pointing AEGIS-3, which has $\sigma$$\sim$0.03 mag), an order of magnitude smaller than the shifts introduced and also comparable or smaller than the uncertainties calculated for the original recalibration offsets (see Table \ref{table:zp_offsets}). 

This suggests that the recalibration procedure converges to the same or very similar stellar population models for most galaxies, regardless of any systematic shifts in the photometry. Therefore, we conclude that the offsets $\Delta m(j)$ obtained in the recalibration remove real systematics in the photometry, probably associated to imperfect aperture corrections. 

We emphasise that the $\Delta m(j)$ in Table \ref{table:zp_offsets} are valid only for the PSFCOR aperture. Also, we warn the reader about the systematic errors that could be introduced if these recalibration offsets are used in the spectral analysis of miniJPAS sources. Their values may still be model-dependent with regards to the specific choice of the stellar library, the extinction law, or other model parameters. 
However, this is not a problem for our photo-$z$ calculation since the template library that we use is a subset of the model grid (see Sect. \ref{subsec:gentemplates}).

\section{Computation of photometric redshifts}

\subsection{Photo-$z$ code}

We compute photo-$z$ for miniJPAS sources using {\sc jphotoz}, a python package that is part of {\sc jype}, the data reduction pipeline for J-PAS. {\sc jphotoz} acts as an interface between the database and the actual photo-$z$ computing code(s) and also handles all the pre- and post-processing of the data, including: application of Galactic extinction correction and recalibration offsets to the photometry, filtering of flagged photometric measurements, contrast-correction of the redshift probability distribution function ($z$PDF; see Sect. \ref{sec:contrast-correction}), and computation of the $odds$ parameter as well as other parameters derived from the $z$PDF (Sect. \ref{sec:scalar-parameters}).

The photo-$z$ code used for miniJPAS is a customised version of {\sc LePhare} \citep{Arnouts11}, modified to remove a limitation in the maximum number of bands in the photometric catalogue (32 in the original {\sc LePhare}), as well as to allow for higher resolution in redshift with a finer sampling of the redshift search range (in miniJPAS, from $z$=0 to $z$=1.5 in constant steps of $\delta z$=0.002). 
{\sc LePhare} computes photo-$z$ using the template-fitting method \citep[see][for a recent review of the different photo-$z$ techniques]{Salvato19}. 
Template fitting photo-$z$ works by evaluating a goodness-of-fit estimator (typically $\chi^2$) between the observed photometry and synthetic photometry generated from each of the templates as a function of $z$.  
The corresponding $\chi^2$ values sample the (assumed Gaussian) log-likelihood distribution, $\log \mathcal{L}(z) \propto -\chi^2_{min}(z)/2$.
Most modern photo-$z$ codes (including {\sc LePhare}) compute the $z$PDF by weighting $\mathcal{L}(z)$ with a redshift prior that summarises our \textit{a priori} knowledge of the underlying redshift distribution as a function of the galaxy magnitude and/or colour (see \citet{Ilbert06} for details and \citet{Benitez00} for a general description of the method).

\begin{figure} 
\begin{center}
\includegraphics[width=8.4cm]{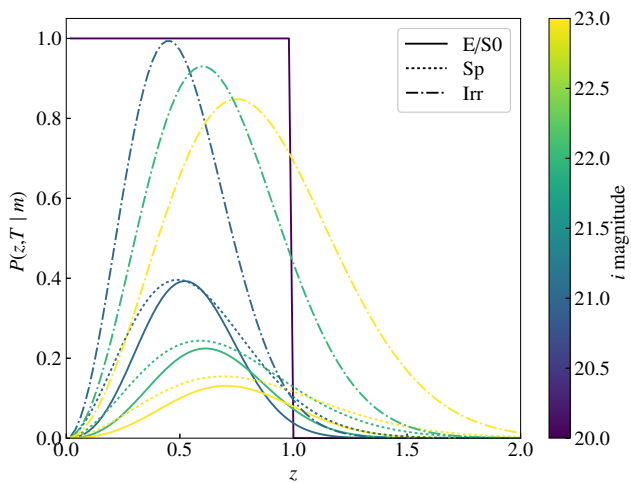}
\end{center}
\caption[]{Prior probability redshift distributions used by {\sc LePhare,} for the broad spectral types ``elliptical'' (solid line), ``spiral'' (dotted line), and ``irregular'' (dot-dashed line), with colour coding for the magnitude (see text for details).\label{fig:priors}}
\end{figure}

We use the default redshift prior in {\sc LePhare}, which is obtained from the spectroscopic redshift distribution of galaxies in the VIMOS VLT Deep Survey \citep[VVDS;][]{LeFevre05}, and contains the probability density function $P$($z$,$T$|$m$) of the redshift and spectral type as a function of the $i$-band magnitude. 
Figure \ref{fig:priors} shows $P$($z$,$T$|$m$) for the broad spectral types ``elliptical'' (E/S0), ``spiral'' (Sp), and ``irregular'' (Irr), defined according to the rest-frame $g$-$i$ colour of the templates (see \citet{Ilbert06} for details), and for magnitudes $i$=21, 22, and 23.
For galaxies brighter than $i$ = 20 the empirical prior is replaced by a step function that takes the value 1 at $z$$\le$1 and 0 at $z$$>$1, independently on the spectral type. 
While such a prior is not a realistic model for the actual redshift distribution of bright galaxies, this fact is unimportant in practice since $i$$<$20 miniJPAS galaxies have high S/N photometry and their $\log \mathcal{L}(z)$ present a very sharp peak.
The prior also prevents $i$$<$20 sources from finding solutions at $z$$>$1. This is not an issue for miniJPAS given its small volume (the brightest galaxy with $z_{\rm{spec}}$$>$1 has $i$=20.6), but a more realistic prior will be needed for J-PAS in order to obtain accurate photo-$z$ for the most luminous $z$$>$1 galaxies.

{\sc LePhare} also calculates separately the minimum $\chi^2$ obtained with a set of stellar and quasar templates, which could help to classify sources. However, we find it more convenient to exclude such templates and to interpret the $z$PDF from {\sc LePhare} as the redshift probability distribution conditional to the source being a galaxy, $P$($z$|G). 

\subsection{{\sc cefca\_minijpas} library of templates}\label{subsec:gentemplates}

\begin{figure} 
\begin{center}
\includegraphics[width=8cm]{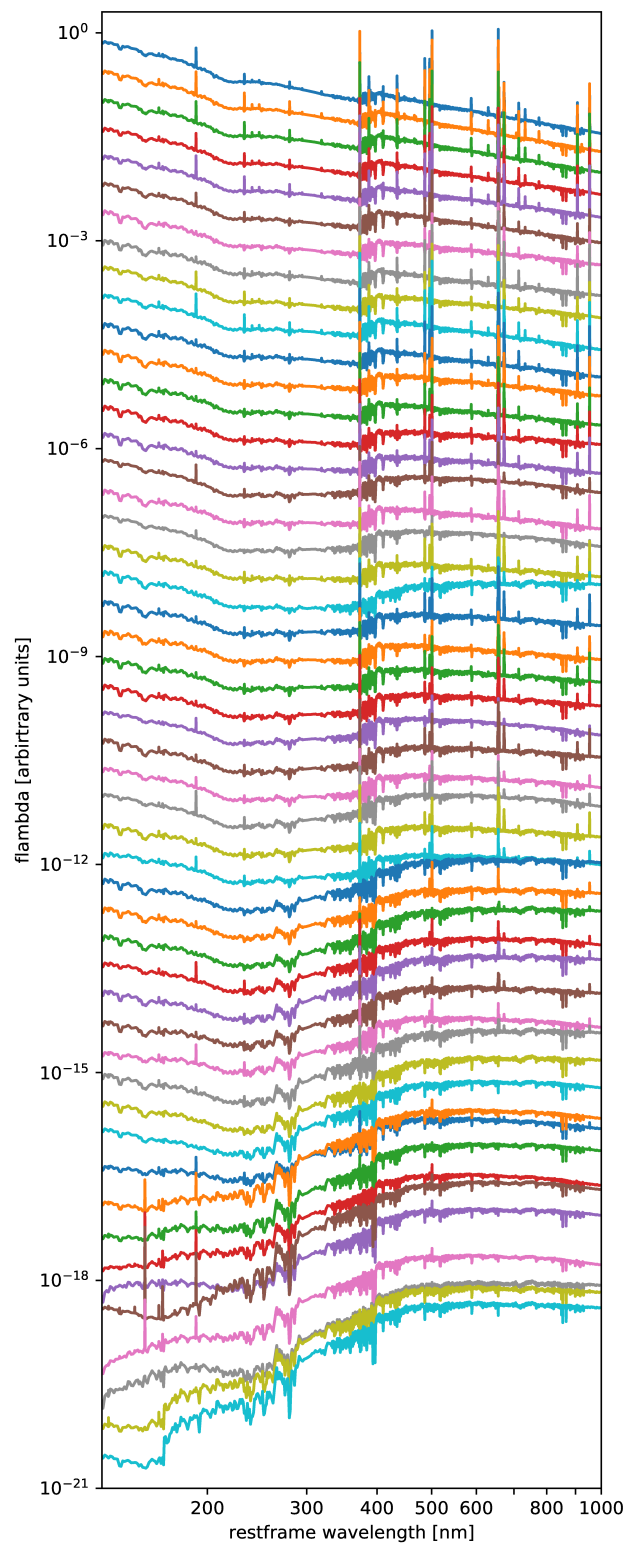}
\end{center}
\caption[]{The 50 galaxy templates used for computation of photo-$z$ for miniJPAS sources, sorted by their rest-frame UV-to-optical spectral index and shifted in the Y axis for clarity.\label{fig:templates}}
\end{figure} 

A key aspect of the template-fitting method for photo-$z$ determination is that generally it does not require a good fit between the template and the observed photometry; it suffices that the best fit is obtained at the true redshift. In particular, 
only a few templates (broadly corresponding to the major spectral types) are needed to obtain good photo-$z$ from broadband photometry \citep[e.g.][]{Benitez00}, where photo-$z$ accuracy depends on the correct detection of broad spectral features like the Lyman break or the 4000 \AA{} break.

The requirements are different when using narrow-band photo-spectra, since narrow spectral features such as emission and absorption lines have a much stronger impact on the photometry (the effect of a spectral feature on the photometry is proportional to its equivalent width and inversely proportional to the bandpass). Multiple works have shown that photo-$z$ for narrow-band datasets improve significantly when emission lines are taken into account \citep[e.g.][]{Molino14,Eriksen19,Alarcon21}

{\sc LePhare} can add emission lines to the templates with intensities individually adjusted for each galaxy to match the photometry in the affected bands. However, this often results in too much freedom, as it allows unphysical models that lead to spurious solutions. Because of this, we prefer to switch off the emission line adjustment capability in {\sc LePhare} and instead provide templates that include both the stellar and nebular emission (lines and continuum).

In the past, template libraries were conditioned by the availability of spectral templates, which were often based on the observed spectra of archetypal galaxies or composites from multiple galaxies of the same type, often extrapolated with models or photometry beyond the range of the spectroscopic observations. 
An alternative approach that has gained popularity thanks to improvements in the stellar libraries and stellar evolution models is the generation of synthetic galaxy spectra with SPS models \citep[e.g.][]{Brammer08,Eriksen19}. 

To build the {\sc cefca\_minijpas} library of templates we start from the grid of 90,720 SPS models that we used for the recalibration. Because we want to use templates that represent real galaxies, we choose only those that provide the best fit for one (or more) of the miniJPAS galaxies using the recalibrated PSFCOR photometry. We exclude sources fainter than $r$ = 22 to guarantee a good S/N in all the bands, and also consider only sources with secure spectroscopic redshifts. We make sure that the best-fitting template is consistent with the observed photometry by requiring a reduced chi-squared $\chi^2_r$ $<$ 1.
We also impose a constraint on the coefficient of variation of the RMS error:

\begin{equation}
CV_{\rm{RMSE}} = \frac{\sqrt{\sum_{n=1}^N \big{(}f_{obs}(\lambda_n) - f_{model}(\lambda_n)\big{)}^2/N}}{\sum_{n=1}^N f_{obs}(\lambda_n)/N}
\end{equation}

\noindent which represents the ratio of the typical residual to the mean flux.
We require CV$_{\rm{RMSE}}<$0.04 because selecting for low $\chi^2_r$ alone favours the sources with lower S/N \citep[see e.g.][]{Hernan-Caballero15}, for which  a wider range of models is consistent with the observed photometry. There are 455 SPS models meeting all these criteria. 

We further reduce the number of templates by iteratively performing photo-$z$ calculation with multiple combinations of these 455 models to select the ones that provide the best results in the spectroscopic subsample. This method for optimisation of the template set is the subject of an upcoming paper (Hern\'an-Caballero et al., in preparation). Very briefly, a score is associated to each combination of templates, which summarises the quality of the photo-$z$ obtained on a test sample with those templates. Starting from random selections of templates, each set of templates is modified by removing some and adding new ones at random (selected from the population of 455 candidates). The score is computed for every new set. Then, those with higher score are taken as the basis for new sets with higher probability. We iterate until the best score among all the sets maxes out and no further increases are obtained in a predetermined number of iterations. We take the set with the highest score as the final selection, and use it to compute the photo-$z$ for the whole miniJPAS sample. 

We repeated this procedure multiple times with different random initial sets. While the final selection changes from one run to another, the score converges to the same maximum value in all the runs, and the photo-$z$ of the whole miniJPAS sample in two different runs are identical for $\sim$98\% of the sources. Some particular templates always get into the final selection, others seem interchangeable, and the large majority never gets selected.

The final set used to compute the published photo-$z$ of miniJPAS contains 50 templates (number for which the precision saturates) and is shown in Fig. \ref{fig:templates}.
The values of the main parameters used to build these models are listed in Table \ref{table:model-properties} and common observables measured on the models 
are presented in Table \ref{table:observables}.

\subsection{Contrast correction of the $z$PDF\label{sec:contrast-correction}}

The probabilistic nature of the $z$PDF implies that for individual sources, it is often impossible to tell whether they are realistic or not, as any $P$($z$|G) that verifies $P$($z_{\rm{spec}}$|G)$>$0 is consistent with the spectroscopic redshift. 
However, when considering large groups of sources, some statistical tests can determine if the $z$PDF are well behaved. One powerful test is the calculation of the fraction of galaxies in which $z_{\rm{spec}}$ falls within a given confidence interval (CI) of the $z$PDF \citep[e.g.][]{Fernandez-Soto02,Dahlen13,Schmidt13}. If the $z$PDF describes the actual redshift probability distribution, we can expect 10\% of galaxies to fall within any 10\% CI, 20\% in a 20\% CI, and so on. Out of the many possible definitions of a CI, the most useful is the highest probability density (HPD) CI, as proposed initially by \citet{Fernandez-Soto02} and illustrated by \citet{Wittman16}.
The HPD CI is the shortest redshift interval (or union of disjoint intervals) that contains a given fraction of the total area under the $z$PDF distribution.
Therefore, it always encloses the main peak of the $z$PDF. 

\begin{figure} 
\begin{center}
\includegraphics[width=8.4cm]{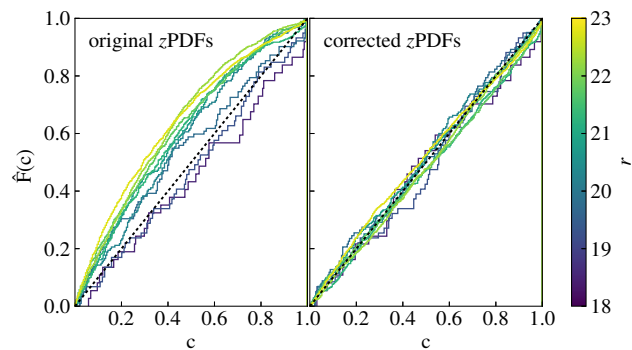}
\end{center}
\caption[]{Fraction of galaxies with $z_{\rm{spec}}$ within the highest probability density confidence interval as a function of the confidence level before (left) and after (right) the contrast correction of the $zPDF$. Galaxies are grouped by their $r$-band flux in bins of 0.5 magnitudes. The diagonal line marks the \^{F}($c$) = $c$ relation expected if the $z$PDF represents the actual redshift uncertainty. Values above (below) this line imply under- (over-)confidence in the $z_{\rm{best}}$ estimate.\label{fig:qqplot}}
\end{figure}

We compute the fraction \^{F}($c$) of miniJPAS galaxies with $z_{\rm{spec}}$ inside the HPD CI at confidence level $c$ using the algorithm presented in \citet{Wittman16}. We separate the galaxies into groups according to their $r$-band magnitude. The results (left panel in Fig. \ref{fig:qqplot}) show that, in general, there are more $z_{\rm{spec}}$ values inside the HPD CI than expected, indicating that the $z$PDF has too much weight at redshifts far from the main peak (or peaks). This effect is increasingly strong at fainter magnitudes. For bright sources ($r$$<$19), the opposite is true: too much weight is placed at the main peak of the $z$PDF, causing \^{F}($c$)$<$$c$.

This type of inaccuracies in the $z$PDF is a common issue of photo-$z$ codes, which causes under- (or over-)estimation of the actual redshift uncertainty  \citep[e.g.][]{Hildebrandt08,Dahlen13,Mundy17}.
Several factors may contribute to these trends, including over- (or under-) estimation of photometric errors and the sparse sampling of the parameter space of the models that results from using only a few templates. The latter implies that the best-fitting template is often substantially different from the actual spectrum of the source.
Such sparseness of the models is particularly relevant for miniJPAS, where the high number of narrow-band filters allows to resolve many spectral features. Because of this, most galaxies get poor fits ($\chi^2_{min}$ $\gg$ 1) even at the spectroscopic redshift, in particular those with higher S/N photometry. 

Modelling the impact on the $z$PDF of all the factors described above is overly complex and ultimately unnecessary. An empirical correction of the contrast of the $z$PDF (the difference between the value at peaks and valleys) suffices to restore the expected relation between \^{F}($c$) and $c$. To this aim, {\sc jphotoz} implements a variation of the method described by \citet{Dahlen13}, which corrects the $z$PDF in two steps.
First, a discrete convolution of the $z$PDF and a Lorentzian kernel is performed:

\begin{equation}
P'[n] = (P * k)[n] = \sum\limits_{m=-2}^{2} P[n-m] k[n] 
\end{equation}
where $P$[$n$] is the value of the $z$PDF for the $n$-th element in which it is sampled (corresponding to the redshift $z$[$n$] = 0.002$n$), and $k$[$m$] = 1/($m^2$ + $\gamma^2$), with $\gamma$=1.18. 
This convolution is particularly important for bright sources, that sometimes have $P$[$n$] = 0 for all elements except the one corresponding to the peak. 
The contrast of the convolved $z$PDF is then adjusted with the transformation $P''$[$n$] = $P'$[$n$]$^{1/\alpha}$, with $\alpha$ = 0.54. 
The values of $\alpha$ and $\gamma$ are calculated by minimising over the whole sample the function:
\begin{equation}
D(\alpha,\gamma) = \sum_i \vert \hat{F}(c_i)-c_i \vert
\end{equation}

The right panel in Fig. \ref{fig:qqplot} shows the relation between \^{F}($c$) and $c$ for the same magnitude cuts after the contrast correction.

\subsection{Scalar parameters\label{sec:scalar-parameters}}

While the $z$PDF provides the most complete description of our knowledge of the redshift of a source, it is often more convenient to use scalar parameters that condense its main properties. The table \textit{PhotoZLephare\_updated} in the miniJPAS database\footnote{http://archive.cefca.es/catalogues/minijpas-pdr201912} contains several of them:
\begin{itemize}
\item[-]{\textbf{Z\_ML} (Z\_ML in {\sc LePhare}) is the median redshift of the $z$PDF (50\% of the total area is on each side).}\\ 

\item[-]{\textbf{PHOTOZ} (Z\_BEST in {\sc LePhare}, hereafter $z_{\rm{best}}$) is the redshift corresponding to the absolute maximum of the $z$PDF. We consider this the most useful point estimate of the photo-$z$, as it is more robust than Z\_ML for asymmetric or multi-peaked $z$PDF profiles.}\\

\item[-]{\textbf{CHI\_BEST} (CHI\_BEST in {\sc LePhare}) is the $\chi^2$ of the best-fitting galaxy model at $z_{\rm{best}}$.}\\

\item[-]{\textbf{Z\_BEST68\_LOW} (Z\_BEST68\_LOW in {\sc LePhare}) the low-$z$ limit of the 68\% confidence interval for $z_{\rm{best}}$, computed using the $\Delta\chi^2$ method \citep[e.g.][]{Anvi76,Bolzonella00}.}\\

\item[-]{\textbf{Z\_BEST68\_HIGH} (Z\_BEST68\_HIGH in {\sc LePhare}) the high-$z$ limit of the 68\% confidence interval for $z_{\rm{best}}$, computed using the $\Delta\chi^2$ method.}\\

\item[-]{\textbf{PHOTOZ\_ERR} is the 1-$\sigma$ uncertainty in $z_{\rm{best}}$, computed as: PHOTOZ\_ERR =  0.5(Z\_BEST68\_HIGH - Z\_BEST68\_LOW). In this paper we use a related quantity, the relative 1-$\sigma$ uncertainty:}

\begin{equation}
z_{err} = \frac{\textrm{Z\_BEST68\_HIGH} - \textrm{Z\_BEST68\_LOW}}{2(1 + \textrm{Z\_BEST})}
\end{equation}

\item[-]{\textbf{ODDS} (hereafter $odds$) is the probability of the relative error in $z_{\rm{best}}$ being smaller than 3\% ($\vert \Delta z \vert$ = $\vert z_{\rm{best}}$-$z_{\rm{spec}} \vert$)/(1+$z_{\rm{spec}}$)$<$0.03). Unlike all the others, this is not a direct output from {\sc LePhare}, but is computed  by {\sc jphotoz} from the contrast-corrected $z$PDF, using the formula:}

\begin{equation}
odds = \int_{z_{\rm{best}} - d}^{z_{\rm{best}} + d} P(z \vert \textrm{G}) dz, \hspace{1cm} d = 0.03(1+z_{\rm{best}})
\end{equation}
 
\end{itemize}

\section{Quantifying the photo-$z$ accuracy}

\subsection{Origin of errors in photo-$z$}

Taking the spectroscopic value as the true redshift of the galaxy, the error in $z_{\rm{best}}$ is often defined as: $\Delta z$ = ($z_{\rm{best}}$ - $z_{\rm{spec}}$)/(1 + $z_{\rm{spec}}$),
where the 1+$z_{\rm{spec}}$ factor conveniently compensates for the stretching of spectral features with redshift. The value of $\Delta z$ is determined by two main types of error, which differ substantially in their prevalence and impact on the resulting photo-$z$. Broadly speaking, we can describe them as ``inaccuracies'' and ``catastrophic errors''.

The former are the cumulative effect of small systematic and random errors in the photometry, flux calibration, and wavelength calibration, as well as the uncertainty introduced by the limited spectral resolution of the photo-spectra, the finite number of templates, and the discretisation of the redshift search range, to name a few. Each of these factors influences the shape and peak redshift of the $z$PDF. Combined, they result in a small, largely random shift in $z_{\rm{best}}$ relative to the true redshift of the galaxy.

On the other hand, catastrophic errors are mainly caused by the non-linearity of the transformation between colour space and redshift, which implies that galaxies with very different spectral types and redshifts may have similar observed colours. The importance of this degeneracy increases at faint magnitudes because larger photometric errors imply more uncertain observed colours which can be consistent with more combinations of template and redshift. The result is a $z$PDF with two or more peaks of comparable strength at different redshifts, often far apart. When the strongest peak is not the one corresponding to the actual redshift, $\Delta z$ can be an order of magnitude or more larger than typical inaccuracies.
The use of redshift priors mitigates this issue, at least in the aggregate, by favouring the most likely redshift given the magnitude of the galaxy. However, for some galaxies with unusually high or low luminosity, the prior may exacerbate the risk of a catastrophic error by favouring the wrong redshift solution (see Sect. \ref{sec:validation-odds}).

Another cause for catastrophic errors is large errors in the photometry that are not accounted for by the nominal flux uncertainties, such as, in contamination by nearby sources or artefacts in the images which may drastically alter the photometry in one or more bands. The result is often spikes or jumps in the photo-spectra that the photo-$z$ code tries to match to legitimate spectral features. Fortunately, in the case of miniJPAS the photometry flags identify most of these sources, and the affected bands are masked for photo-$z$ computation. However, a small number of galaxies is likely to be affected by yet undetected issues.
Finally, galaxies with ``exotic'' spectra that do not resemble any of the galaxy templates in the library may also get wrong redshift estimates.

\begin{figure} 
\begin{center}
\includegraphics[width=8.4cm]{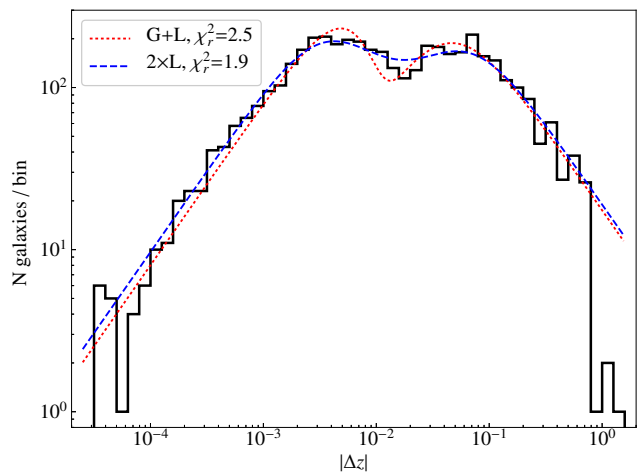}
\end{center}
\caption[]{Distribution of redshift errors for the spectroscopic subsample of miniJPAS (histogram) and best-fitting model combining one Gaussian and one Lorentzian profile (dotted line) or two Lorentzians (dashed line).\label{fig:dz-distrib-model}}
\end{figure}

Figure \ref{fig:dz-distrib-model} shows the distribution of $\vert\Delta z\vert$ for the miniJPAS galaxies with $r$$<$23 and with spectroscopic redshifts. 
The binning is uniform in $\log$ $\vert\Delta z\vert$ to highlight the bimodality of the distribution caused by the two types of error. The main peak at $\vert\Delta z\vert$$\sim$0.004 corresponds to the typical error due to inaccuracies, while the peak at $\vert\Delta z\vert$$\sim$0.04 represents the catastrophic errors. 
The first peak and the tail at very small $\vert\Delta z\vert$ are well reproduced by a Gaussian or a Lorentzian distribution. The second peak requires a Lorentzian, since the slope of the Gaussian at large $\vert\Delta z\vert$ is too steep. Also, the combination of two Lorentzians fits the distribution of $\vert\Delta z\vert$ slightly better than the combination Gaussian+Lorentzian.

As we will see in the next sections, the relative importance of inaccuracies and catastrophic errors in shaping the distribution of $\Delta z$ varies with properties of the galaxies such as the brightness and redshift, and can be predicted (to some extent) from parameters derived from the $z$PDF such as $z_{err}$ and $odds$.

\subsection{Summary statistics}

Similarly to the scalar parameters derived from the $z$PDF, it is often very convenient to rely on summary statistics of the distribution of $\Delta z$ in the analysis of the dependence of photo-$z$ accuracy with one or more galaxy properties, or for easy comparison between samples. 
Since, in most cases, the distribution of $\Delta z$ is far from Gaussian, the mean and standard deviation are often replaced by more ``robust'' analogs: the median and the normalised median absolute deviation, \snmad. The latter is defined as

\begin{equation}
\sigma_{\rm{NMAD}} = 1.48 \times \textrm{median}\big{\vert}\Delta z_i - \textrm{median}(\Delta z_i)\big{\vert}
\end{equation}
\noindent where the factor 1.48 is used to match the standard deviation for a Gaussian distribution. Like the standard deviation, $\sigma_{\rm{NMAD}}$ is insensitive to a systematic offset in $z_{\rm{best}}$, but unlike the former, it is also insensitive to the tail of the distribution for large values of $\vert \Delta z \vert$.

A complementary statistic to $\sigma_{\rm{NMAD}}$ is the outlier rate, $\eta$, which represents the fraction of galaxies with redshift errors larger than a given threshold $X$:

\begin{equation}
\eta = \frac{N(\vert \Delta z \vert > X)}{N_{tot}}
\end{equation}

\noindent where $X$ is set at several times the $\sigma_{\rm{NMAD}}$ (in our case, we choose $X$ = 0.03) to ensure that only values far from the main peak of the distribution are identified as outliers. As a consequence, $\eta$ is a good approximation for the frequency of catastrophic errors.  

While $\sigma_{\rm{NMAD}}$ and $\eta$ are aggregate statistics that describe the  redshift errors of samples, not individual galaxies, their values can also help validate uncertainty estimates for individual sources. In particular, if redshift uncertainties obtained with the $\Delta\chi^2$ method are accurate, we can expect $\sigma_{\rm{NMAD}}$ $\sim$ $\langle$$z_{err}$$\rangle$ for a sample of galaxies selected to have similar values of $z_{err}$. Also, in a sample of galaxies with comparable $odds$, the expected outlier rate is $\eta$ $\sim$ 1 - $\langle$$ odds$$\rangle$. In Sects. \ref{sec:validation-zerr} and \ref{sec:validation-odds}, we apply these tests to the miniJPAS sample.

\section{Results\label{sec:results}}

We compute photo-$z$ for nearly all $r$$<$24 sources in the dual mode photometric catalogue of miniJPAS. However, we will restrict the analysis to $r$$<$23 sources, as fainter ones are undetected in most or all of the narrow-band images, and the completeness of miniJPAS for extended sources also drops quickly at $r$$>$23 (see Fig. 16 in \citetalias{Bonoli21}).

Out of 20,962 miniJPAS sources with $r$$<$23, 186 (0.9\%) were not selected for photo-$z$ calculation due to not meeting the $FLAGS$$<$4 condition in the selection band (these sources have PHOTOZ = -1 in the redshift catalogue). For another 87 (0.4\%) {\sc LePhare} could not obtain a photo-$z$ measurement due to non-detections in all but one non-flagged bands (PHOTOZ = -99 in the catalogue).

Additionally, we consider as invalid the redshift solutions for all sources where $z_{\rm{best}}$ is at one of the extremes of the redshift search range ($z_{\rm{best}}$ = 0 or $z_{\rm{best}}$ = 1.5). This is because we cannot determine whether a minimum of $\chi^2$($z$) found at one of the extremes is a local minimum, and anyway the actual number of miniJPAS galaxies at $z$=0 or $z$=1.5 must be very small.
Out of 203 sources with $z_{\rm{best}}$ = 0, 200 are clearly stars ($P_S$$>$0.99) while the remaining 3 are spurious sources in the halo of a bright star. 
40 out of 69 sources with $z_{\rm{best}}$ = 1.5 are also point sources, most of them known quasars.

The final number of miniJPAS sources with $r$$<$23 and valid photo-$z$ is 20,417. Out of these, 15,719 have FLAGS=0 in all bands. In this section we use the later to characterise the photo-$z$ of miniJPAS.

\subsection{Redshift distribution of miniJPAS galaxies}

\begin{figure} 
\begin{center}
\includegraphics[width=8.4cm]{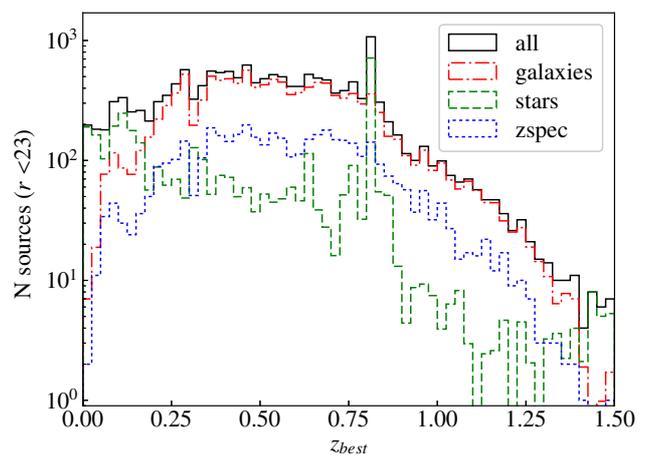}
\end{center}
\caption[]{Distribution of the most likely redshift ($z_{\rm{best}}$) for all miniJPAS sources brighter than $r$=23 (black solid line), as well as for galaxies (red dot-dashed line), stars (green dashed line), and galaxies with spectroscopic redshifts (blue dotted line).\label{fig:Nz}}
\end{figure}

Figure \ref{fig:Nz} shows the distribution of $z_{\rm{best}}$ for miniJPAS sources brighter than $r$$<$23. The black histogram is obtained by assigning the same weight $w$ = 1 to every source regardless of its morphology, while the green and red ones weight each source by the probability of being a star ($w$ = $P_S$) or a galaxy ($w$ = $P_G$ = 1 - $P_S$), respectively.

While all stars are obviously at $z$=0, their $z_{\rm{best}}$ estimates span the entire 0$<$$z$$<$1.5 search range. Stars dominate the number counts at the extremes ($z_{\rm{best}}$$<$0.2 and $z_{\rm{best}}$$>$1.4) due to a lower number of galaxies detected at these redshifts. There is also a very strong peak at $z_{\rm{best}}$$\sim$0.82 that is caused by M-type stars, which find their best fit at that particular redshift. Other weaker peaks in the distribution of $z_{\rm{best}}$ for stars are also evident in Fig. \ref{fig:Nz}.

Since galaxies represent $\sim$80\% of the $r<$23 sample, their redshift distribution is very similar to that for all sources, except at both ends of the redshift range where galaxy counts fall steeply. 
The distribution of $z_{\rm{best}}$ for galaxies is not smooth, but changes from one redshift bin to the next by more than expected from shot noise.
Interestingly, these peaks are mirrored in the distribution of $z_{\rm{spec}}$, indicating that miniJPAS can trace over-densities and voids in the radial direction at least up to $z$$\sim$0.8. 
It is unclear if structure in $z_{\rm{best}}$ at $z$$\gtrsim$1 also corresponds to real over/under-densities, since the spectroscopic counts are too low in this range. In any case, the general trend remains consistent between $z_{\rm{best}}$ and $z_{\rm{spec}}$ within the uncertainties up to the photo-$z$ search range limit of $z$=1.5.

\begin{figure} 
\begin{center}
\includegraphics[width=8.4cm]{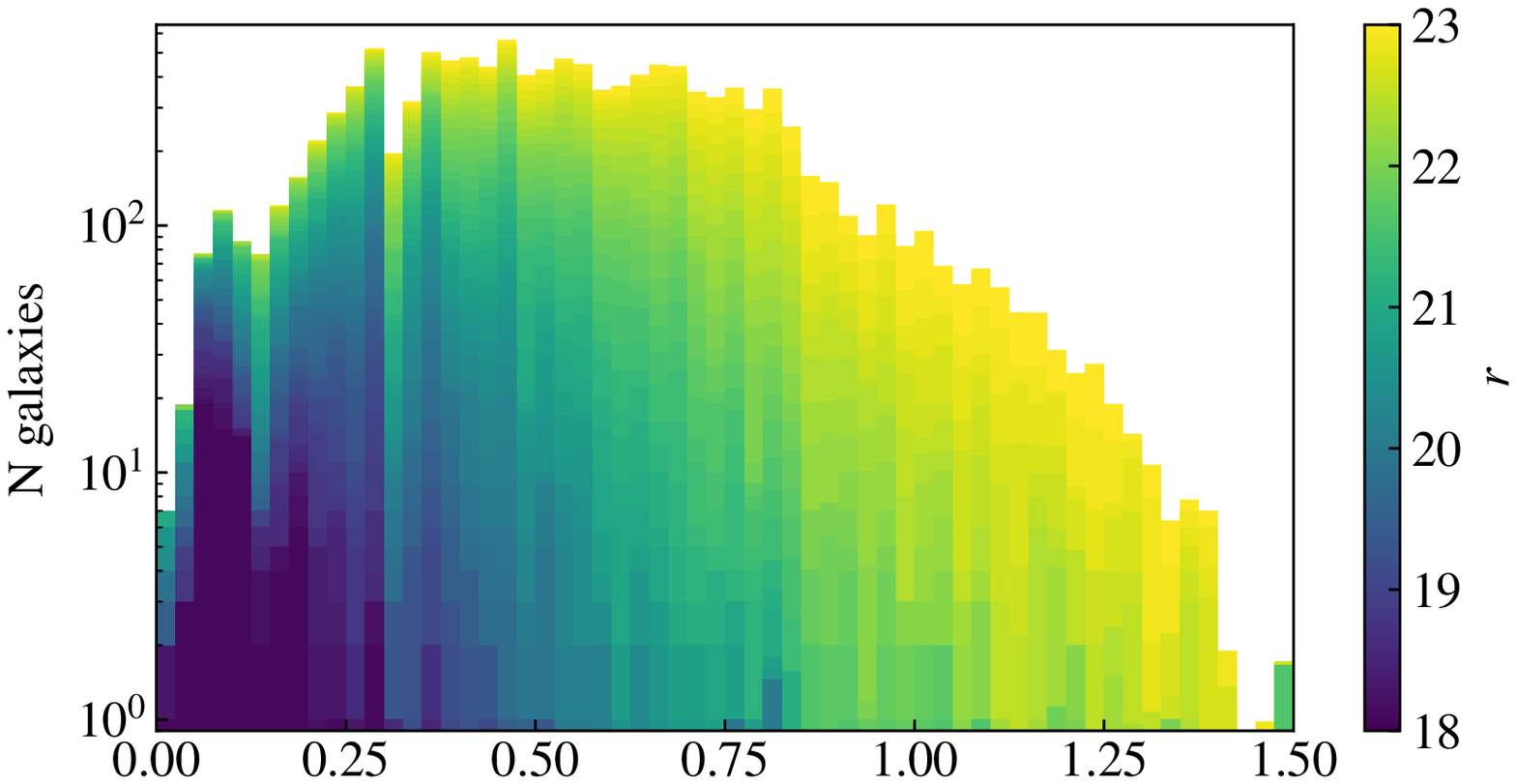}
\includegraphics[width=8.4cm]{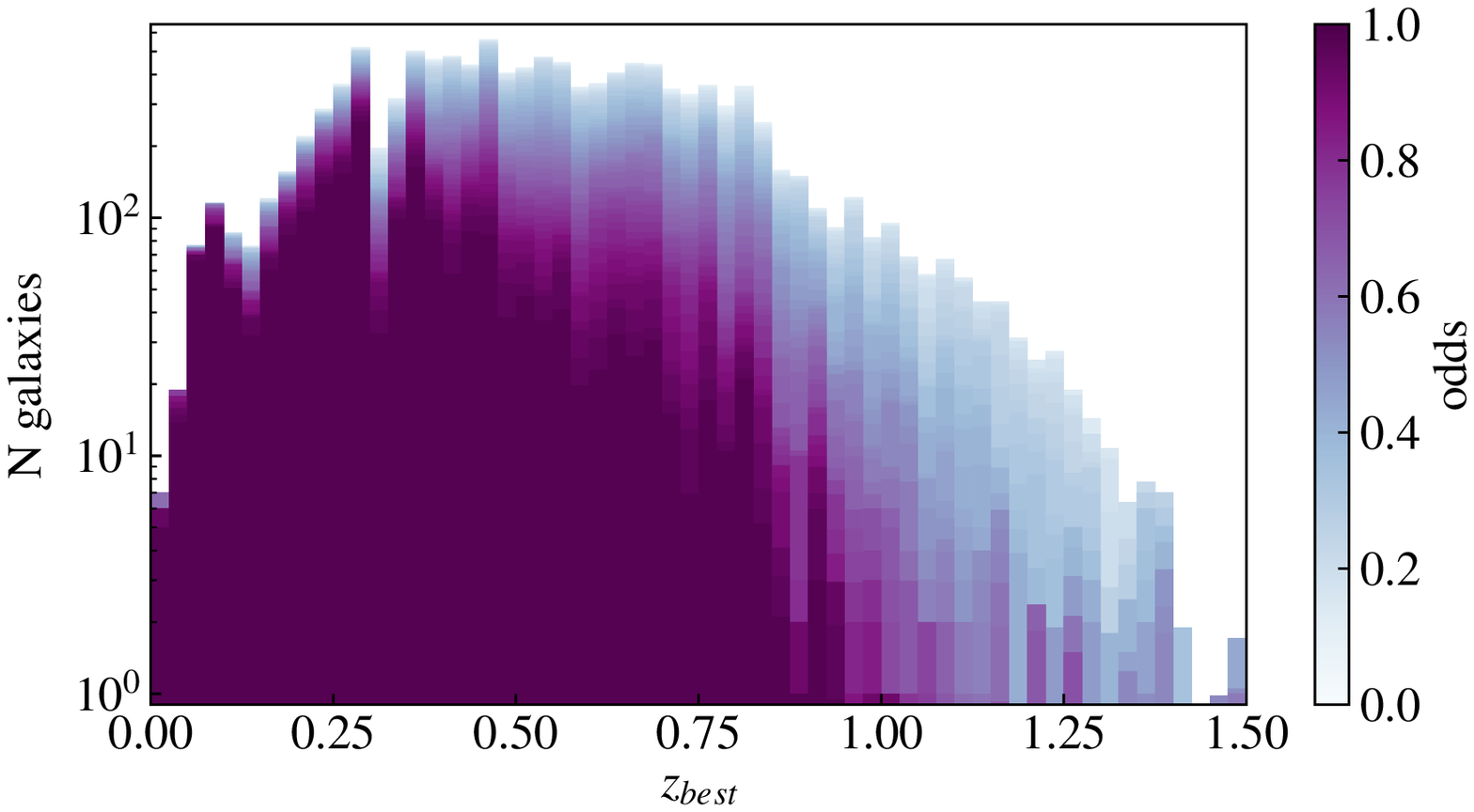}
\end{center}
\caption[]{Redshift distribution of miniJPAS galaxies as a function of the magnitude cut (top panel) or odds cut (bottom panel) applied on the sample.\label{fig:Nzcolormap}}
\end{figure}

\begin{figure} 
\begin{center}
\includegraphics[width=8.4cm]{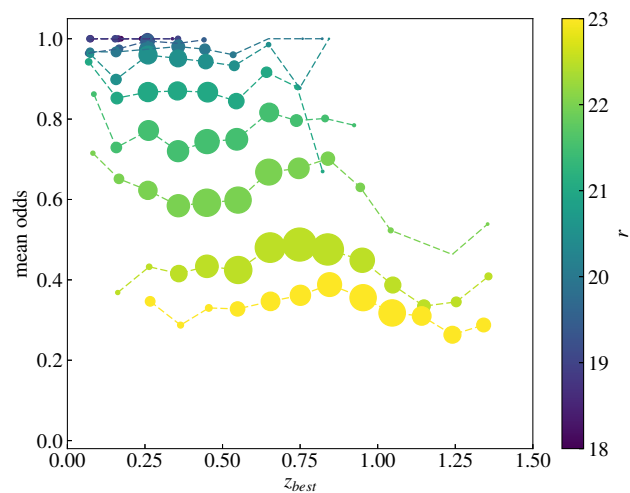}
\end{center}
\caption[]{Mean value of the odds parameter for galaxies as a function of redshift and brightness. Each symbol indicates the average odds for galaxies in bins of width equal to 0.1 in $z_{\rm{best}}$ and 0.5 in $r$-band magnitude. Symbol areas are proportional to the effective number of galaxies contributing to each bin.\label{fig:mean-odds}}
\end{figure}

\begin{figure*} 
\begin{center}
\includegraphics[width=17cm]{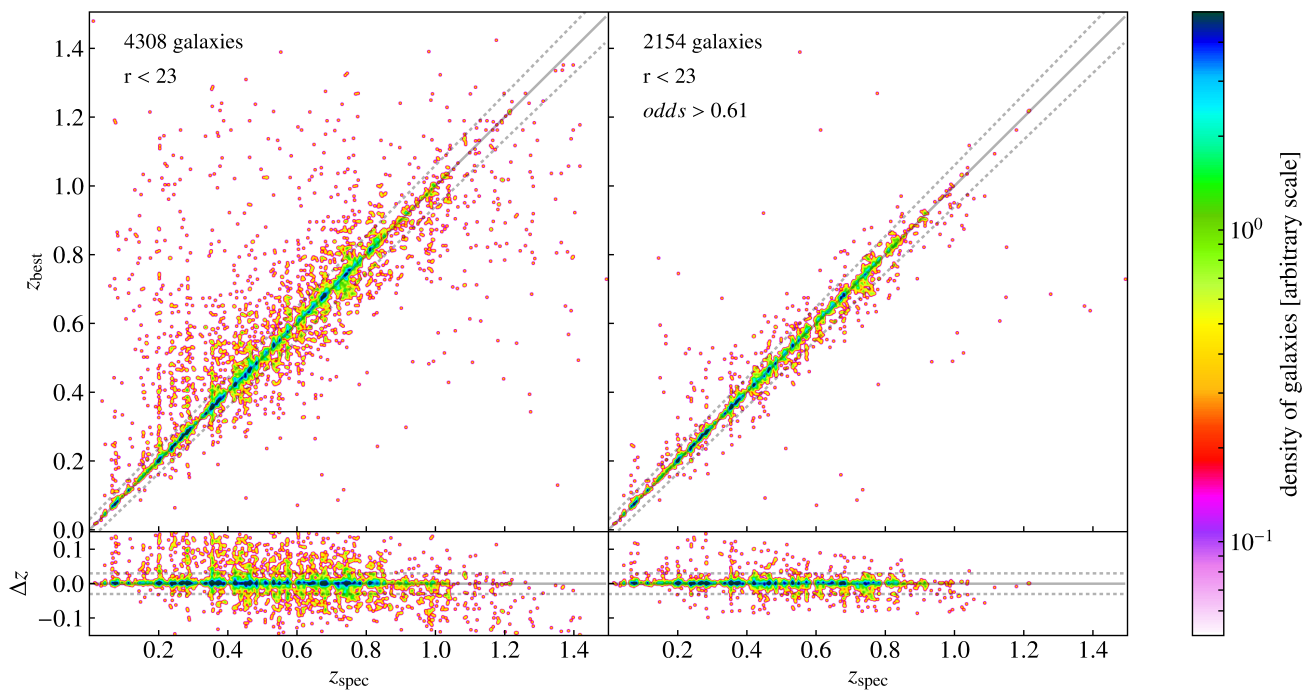}
\end{center}
\caption[]{Comparison between photometric and spectroscopic redshifts for individual miniJPAS galaxies in the spectroscopic sample. The left panel includes all $r$$<$23 galaxies with valid photo-$z$ estimates while the right one contains only half the sample (those with higher $odds$). The bottom panels show the redshift errors, $\Delta z$. A 2-D Gaussian smoothing is applied to the data to improve the visualisation of the density of points. The solid line marks the $z_{\rm{best}}$ = $z_{\rm{spec}}$ relation, while the dotted lines indicate the $\vert \Delta z \vert$ = 0.03 threshold used to define outliers.\label{fig:zphot-zspec-map}}
\end{figure*}

The redshift distribution depends strongly on the magnitude of the sources, since higher redshift galaxies are typically fainter. The top panel in Fig. \ref{fig:Nzcolormap} shows the distribution of $z_{\rm{best}}$ as a function of the limiting $r$-band magnitude of the selection. Nearly all $z_{\rm{best}}$$>$1 galaxies are faint ($r$ $\gtrsim$ 22), implying low S/N in the narrow bands. This makes their $z_{\rm{best}}$ estimates unreliable (low $odds$, see bottom panel in Fig. \ref{fig:Nzcolormap}). 
However, the decrease in the $odds$ at high redshift is steeper than expected from the increase in the average magnitude alone. 

Figure \ref{fig:mean-odds} shows that at constant magnitude, the mean $odds$ actually increases slightly from $z$$\sim$0.5 to $z$$\sim$0.8 but decreases abruptly at $z$$>$0.8. This is probably a consequence of the highest contrast spectral features (in particular H$\alpha$ and the 4000 \AA{} break) shifting into redder bands (where the depth is shallower, see Fig. 4 in \citetalias{Bonoli21}) and ultimately out of the miniJPAS range.

\subsection{Accuracy of $z_{\rm{best}}$ estimates\label{sec:errors}}

In this section, we use the subsample of miniJPAS galaxies with spectroscopic redshifts to evaluate the accuracy of photo-$z$ determinations using the most probable value (the mode of the $z$PDF), $z_{\rm{best}}$. 
Figure \ref{fig:zphot-zspec-map} shows the usual comparison between photometric and spectroscopic redshifts. In the case of miniJPAS, a normal scatter plot is not very informative since most of the dots clump in a very narrow band around the diagonal line that marks the 1:1 relation. To give a realistic impression of the actual density of dots along this line, we generate a density map from the scatter plot by convolving with a Gaussian kernel (note that the pixelation and convolution performed to generate the density map cause some broadening of the distribution compared to the original scatter plot).

The dark blue areas indicate the regions with highest density of dots. These correspond to the over-densities found in the distribution of $z_{\rm{spec}}$ in Fig. \ref{fig:Nz}. The dotted lines enclosing the $\vert \Delta z \vert$ $<$ 0.03 region contain 64\% of the whole $r$$<$23 sample (left panel) and 87\% of the subsample with $odds$$>$0.61 (right panel).

Comparison of the left and right panels in Fig. \ref{fig:Nz} shows that most of the dispersion in the $z_{\rm{best}}$ vs $z_{\rm{spec}}$ relation is due to galaxies with low $odds$. At faint magnitudes the shape of the $z$PDF is increasingly dominated by the redshift prior which favours $z_{\rm{best}}$ values close to the peak probability defined by the prior at each magnitude (see Fig. \ref{fig:priors}).
This implies large errors in $z_{\rm{best}}$ for sources whose true redshift is far from the $z$$\sim$0.7 peak of the prior.

\begin{figure} 
\begin{center}
\includegraphics[width=8.4cm]{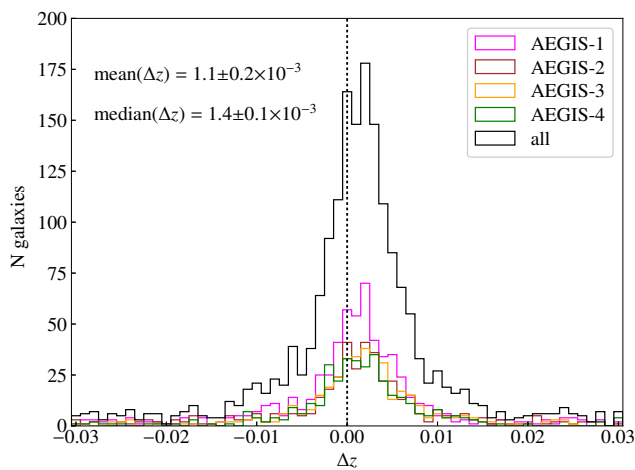}
\end{center}
\caption[]{Distribution of the error in $z_{\rm{best}}$, $\Delta z$, for the galaxies in the spectroscopic sample. The tails of the distribution at $\vert\Delta z\vert$$>$0.03 are truncated to emphasise the shape of the central peak. Only sources with $odds$$>$0.61 are shown. Each colour represents the distribution for an individual pointing. The black histogram represents the combined distribution for all four pointings. \label{fig:dz-hist}}
\end{figure}

\begin{figure} 
\begin{center}
\includegraphics[width=8.4cm]{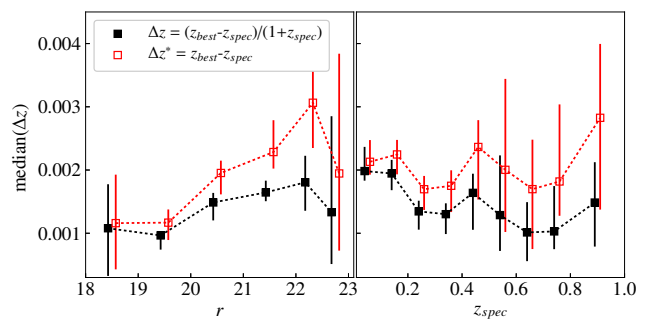}
\end{center}
\caption[]{Median error in $z_{\rm{best}}$ as a function of the $r$-band magnitude (left) and redshift (right) for the galaxies in the spectroscopic sample verifying $\vert \Delta z\vert$$<$0.03 and $odds$$>$0.65. Solid black symbols and open red symbols represent two different definitions of $\Delta z$ (see text for details). Error bars indicate the 16--84$^{th}$ percentile confidence interval obtained with bootstrap resampling.\label{fig:dz-mag-z}}
\end{figure}

The distribution of $\Delta z$ is noticeably shifted from the origin (Fig. \ref{fig:dz-hist}), indicating a small positive systematic bias in $z_{\rm{best}}$. The shift is evident in all the four AEGIS pointings when considered separately.

To estimate the magnitude of the systematic bias in $z_{\rm{best}}$, $\Delta z_{\rm{sys}}$, we calculate the median $\Delta z$ among the galaxies with $odds$$>$0.65 and $\vert \Delta z\vert$$<$0.03. These constraints help to decrease the dispersion introduced by outliers and galaxies with a broad $z$PDF.
We obtain $\Delta z_{\rm{sys}}$ = 1.4$\pm$0.1$\times$10$^{-3}$. The shift is also detected at high significance for other cuts in $odds$ or $r$-band magnitude and for no cuts at all.  

We find no dependence of $\Delta z_{\rm{sys}}$ with the spectral type of the galaxies (see Sect. \ref{sec:spectral-type} for the details of our classification method). 
We also check for a dependence of $\Delta z_{\rm{sys}}$ with the $r$-band magnitude and redshift of the galaxies in Fig. \ref{fig:dz-mag-z}. We find tentative evidence for an increase in the median $\Delta z$ with $r$, and a decrease with $z_{\rm{spec}}$. This is striking, since $r$ and $z_{\rm{spec}}$ have a positive correlation.
The decrease from $\Delta z_{\rm{sys}}$$\sim$0.002 at $z$$\sim$0 to $\Delta z_{\rm{sys}}$$\sim$0.001 at $z$$\sim$0.8 is consistent with a constant offset in $z_{\rm{best}}$ instead of $z_{\rm{best}}$/(1+$z_{\rm{spec}}$). To prove this, we show in red open symbols the median value of $\Delta z^*$ = $z_{\rm{best}}$ - $z_{\rm{spec}}$, which revolves around $\sim$0.002 for the entire redshift range. While an offset of $\sim$0.002 matches the redshift step used in the photo-$z$ calculation and the $z$PDF, it is unlikely that $\Delta z_{\rm{sys}}$ is related to the discretisation of the redshift range since we find comparable values for a redshift step of 0.001.

\begin{figure} 
\begin{center}
\includegraphics[width=8.4cm]{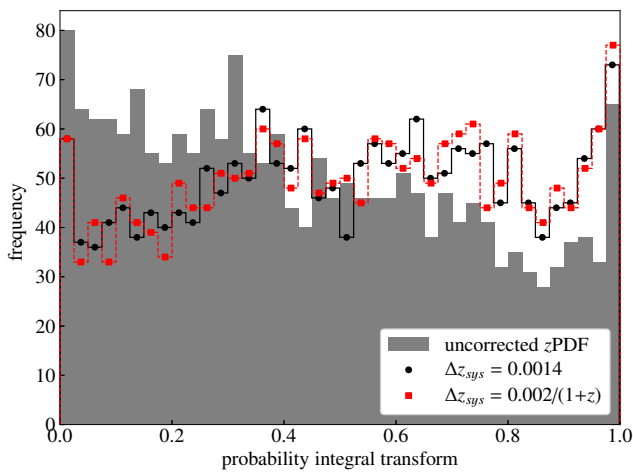}
\end{center}
\caption[]{Distribution of probability integral transform (PIT) values for the $z$PDFs of galaxies in the spectroscopic sample with $odds$$>$0.65. Black circles and red squares represent the PIT distributions after correcting the individual $z$PDFs assuming a systematic offset of $\Delta z_{\rm{sys}}$$\sim$0.0014 and $\Delta z_{\rm{sys}}$$\sim$0.002/(1+$z$), respectively. Grey bars show the PIT distribution with no correction for systematic offset applied.\label{fig:PIT-diagram}}
\end{figure}

The systematic bias affects not only the $z_{\rm{best}}$ values but the entire $z$PDFs, as evidenced by the slope of the distribution of the probability integral transform (PIT) of the $z$PDFs (Fig. \ref{fig:PIT-diagram}). PIT values for individual sources are computed as the cumulative distribution function of the $z$PDF evaluated at the spectroscopic redshift \citep[e.g.][]{Schmidt20} and represent the probability P($z$$<$$z_{\rm{spec}}$). Well calibrated $z$PDFs result in a flat distribution of PIT values, while a tilted distribution indicates a bias \citep{Polsterer16}. Shifting in redshift the $z$PDFs by $\Delta z_{\rm{sys}}$$\sim$0.0014 or $\Delta z_{\rm{sys}}$$\sim$0.002/(1+$z$) largely removes this bias. The excess frequency in the first and last bins of the PIT distribution are the consequence of catastrophic redshift errors \citep{Schmidt20}.

The origin of this bias is uncertain. Systematic errors in the wavelength calibration of the JPAS filters is one possibility, as well as errors in the characterisation of the spectral response of the detector or the telescope throughput. However, to produce a shift of $\Delta z_{\rm{sys}}$ = 1.4$\times$10$^{-3}$ the effective wavelength of the filters would have to be redshifted by $\sim$9 $\AA${} on average, which is more than an order of magnitude larger than the precision of the transmission curves. 
The same bias could be obtained if the spectral templates are blue-shifted by a similar amount, but this is also highly unlikely as the main emission lines are all found at the expected wavelengths. 
The redshift prior can, in principle, bias $z_{\rm{best}}$ values if the width of the prior is comparable to the width of the $zPDF$. However, our redshift prior is flat for $r$$<$20 galaxies and broad for fainter ones, and the bias is also found for sources with high odds that typically have very narrow $z$PDFs. Finally, the existence of a bug in the code of {\sc LePhare} is conceivable, but this possibility has been ruled out since roughly the same $\Delta z_{\rm{sys}}$ is obtained independently with the {\sc TOPz} code (Laur et al., in prep.).
Because $\Delta z_{\rm{sys}}$ is significantly smaller than the nominal uncertainty of most $z_{\rm{best}}$ estimates and its origin remains unknown, we choose not to correct $z_{\rm{best}}$ values for this bias.

\subsection{Best predictor of photo-$z$ accuracy} 

\begin{figure} 
\begin{center}
\includegraphics[width=8.4cm]{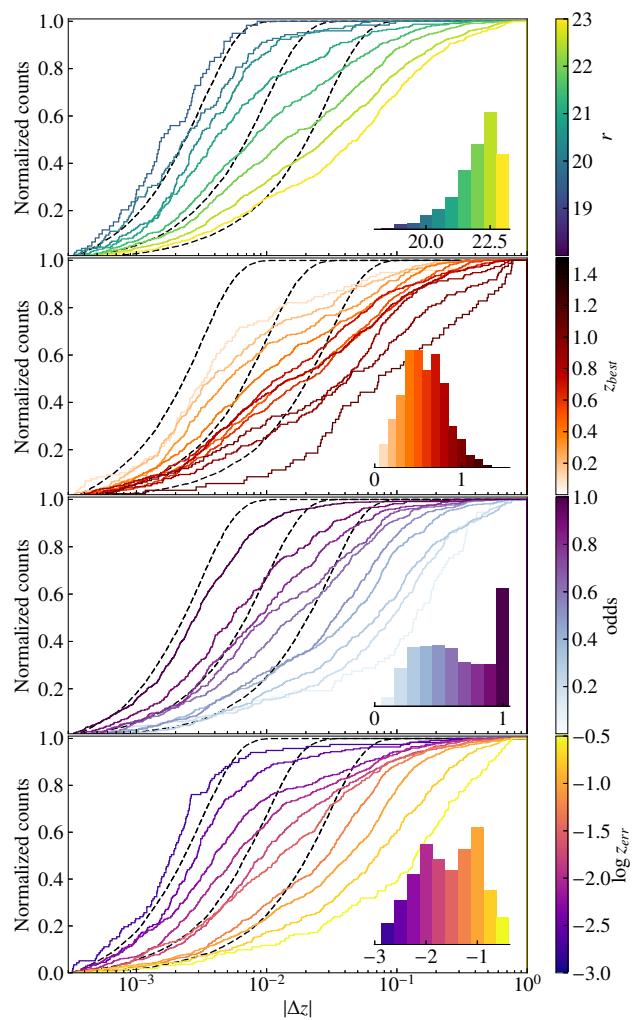}
\end{center}
\caption[]{Cumulative distributions of $\vert\Delta z\vert$ for subsets of the  spectroscopic sample within specific intervals of (from top to bottom) the $r$-band magnitude, $z_{\rm{best}}$, $odds$, and $z_{err}$. Dashed lines represent predictions for a Gaussian distribution of $\Delta z$ centred at $\Delta z_{\rm{sys}}$ = 0.001 and with standard deviations of 0.003, 0.01, and 0.03, while solid lines show the actual distributions for galaxies inside each bin.
The inset plots on the right of the panels represent the relative number of galaxies contributing to each bin, coloured according to the central value of the bin. Cumulative distributions for bins containing less than 50 galaxies are not shown.\label{fig:absdz}}
\end{figure}

The distributions of $\vert \Delta z \vert$ and $\Delta z$ shown in Figs. \ref{fig:dz-distrib-model} and \ref{fig:dz-hist} do not reflect the dependence that photo-$z$ accuracy has on many galaxy properties, such as the magnitude of the sources (which determines the S/N of the photometry), the redshift (which conditions the spectral features within the observed spectral range), and the spectral type (which determines the strength of the spectral features). 
In addition, some parameters derived from the $z$PDF such as $z_{err}$ and $odds$ are not real galaxy properties, but clearly depend on them and, in practice, summarise our knowledge about the multiple factors that impact photo-$z$ performance. All these quantities correlate, to some extent, with each other and with $\vert \Delta z\vert$. 
In this section, we check how the distribution of errors in $z_{\rm{best}}$ depends on the $r$-band magnitude, $z_{\rm{best}}$, $z_{err}$, and $odds$, and analyse which quantity is most useful as a predictor of the photo-$z$ accuracy for individual sources.

Figure \ref{fig:absdz} shows the cumulative distribution of $\vert \Delta z \vert$ for subsamples of the spectroscopic sample, selected according to the value of one of these four quantities. We split the range of variation for each of them into same-width intervals. The relative number of galaxies in every interval is shown in the small bar histograms. 
For reference, we plot in dashed lines the cumulative distributions of $\vert \Delta z \vert$ that would result from Gaussian errors with standard deviations of $\sigma$($\Delta z$) = 0.003, 0.01, and 0.03. 

All four quantities present the same general trend: in the most favourable cases (left-most lines, corresponding respectively to bright $r$-band magnitude, low $z$, high $odds$, or low $z_{err}$), the distributions of $\vert \Delta z \vert$ are nearly Gaussian, with departure from Gaussianity only at the high $\vert \Delta z \vert$ tail. However, as we explore less favourable values of the quantities, the distributions shift to higher $\vert \Delta z \vert$ at any given value of the normalised counts. They also become less Gaussian, with flatter slopes and heavier tails at high $\vert \Delta z \vert$. 
The reason for this change in the shape of the $\vert \Delta z \vert$ distributions is an increase in the rate of catastrophic errors as we progress from left to right. 

\begin{figure} 
\begin{center}
\includegraphics[width=8.4cm]{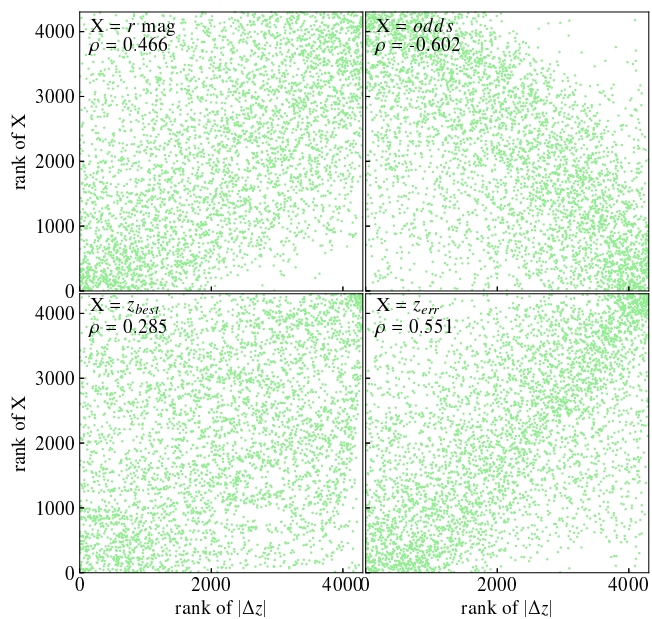}
\end{center}
\caption[]{Comparison of the predictive power for the error in $z_{\rm{best}}$ of four quantities: the $r$-band magnitude, the value of $z_{\rm{best}}$, the $odds$ parameter, and $z_{err}$. Each plot shows the correlation between the ranks of individual galaxies when sorted by each of this quantities and the rank when sorted by $\vert \Delta z \vert$. The Spearman rank correlation coefficient ($\rho$) is quoted for each case.\label{fig:absdz-correlations}}
\end{figure}

To identify which of these four quantities is more effective in separating the good from the bad photo-$z$, we cannot rely on Fig. \ref{fig:absdz} since the distribution of each of these quantities in the miniJPAS sample is different. Instead, we calculate the correlation of these quantities with $\vert \Delta z \vert$ for individual sources (Fig. \ref{fig:absdz-correlations}).
Because of the different ranges that each of these quantities span and the non-linearity of their relation with $\vert \Delta z \vert$, we compute correlations between ranks, instead of correlation of their values. That is, for each quantity (including $\vert \Delta z \vert$) we sort the galaxies in ascending order and compute the Spearman rank correlation coefficient ($\rho$). We find $\rho$ = 0.285, 0.466, 0.551, and -0.602 for $z_{\rm{best}}$, $r$-mag, $z_{err}$, and $odds$, respectively. In the case of the $odds$ parameter, $\rho$ is negative because the average $\vert \Delta z \vert$ decreases at higher $odds$.
According to this, the $odds$ parameter has the strongest correlation with $\vert \Delta z \vert$, closely followed by $z_{err}$, then $r$ while $z_{\rm{best}}$ has the weakest correlation.

This suggests that the best way to select a subsample of $N$ galaxies with the most accurate photo-$z$ is not to pick the $N$ brightest galaxies in the sample (a cut in magnitude) or the $N$ with smallest $z_{err}$, but the $N$ with higher $odds$. The high dispersion in all panels of Fig. \ref{fig:absdz-correlations} implies that for individual galaxies or small samples that may not always be the case, but for large enough $N$, $odds$ should have a clear advantage. 

\begin{figure} 
\begin{center}
\includegraphics[width=8.4cm]{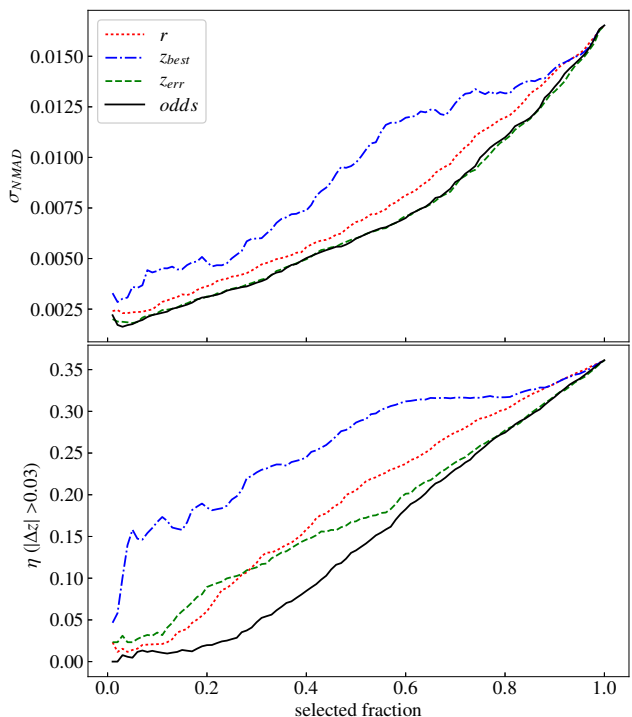}
\end{center}
\caption[]{Dependence of $\sigma_{\rm{NMAD}}$ and $\eta$ on the fraction of the $r$$<$23 sample selected using a threshold in any of four different quantities: $r$-band magnitude, $z_{\rm{best}}$, $z_{err}$, and $odds$. \label{fig:snmad-outrate-completeness}}
\end{figure}

We confirm this hypothesis by computing \snmad{} and $\eta$ as a function of the fraction of the spectroscopic sample selected ($f$) with a threshold in any of these four quantities (Fig. \ref{fig:snmad-outrate-completeness}). 
For $f$ close to 1 (very few galaxies rejected), all four quantities yield almost the same \snmad{} and $\eta$, since the samples selected are also nearly identical. However, as $f$ decreases due to more restrictive thresholds, the tracks diverge.
If we target a specific \snmad, the size of the sample selected with a threshold in $z_{err}$ or $odds$ is significantly larger compared to a selection in $r$-band magnitude (up to $\sim$50\% larger for \snmad$\sim$0.003). If, on the other hand, we target a specific sample size, the \snmad{} obtained from the corresponding threshold in $z_{err}$ or $odds$ is also significantly smaller.

\begin{figure} 
\begin{center}
\includegraphics[width=8.4cm]{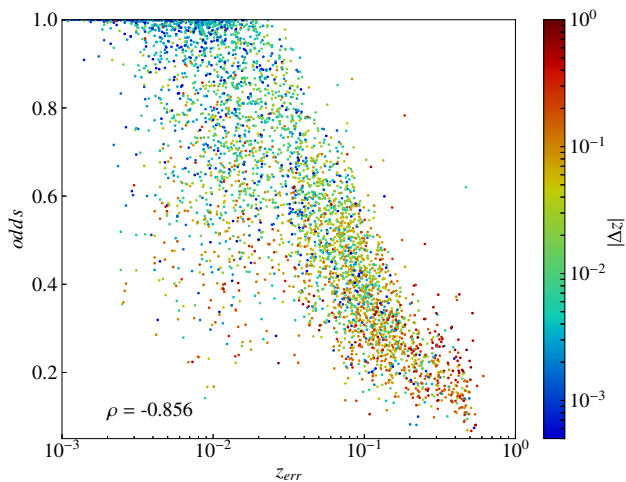}
\end{center}
\caption[]{Correlation between $z_{err}$ and $odds$ for individual sources in the spectroscopic sample.\label{fig:zerr-odds-correlation}}
\end{figure}

The curves of \snmad($f$) for $z_{err}$ and $odds$ are nearly identical. This is no surprise, as the two parameters have a very strong anti-correlation ($\rho$=-0.856, Fig. \ref{fig:zerr-odds-correlation}), much stronger than any of them has with $\vert\Delta z\vert$.  

It is remarkable that $z_{err}$ and $odds$ obtain the same \snmad{} in the whole range of $f$ and that it is significantly better than that of the $r$-band magnitude. However, $z_{err}$ has much worse performance when it comes to avoiding outliers, in particular at $f$$<$0.4 where it is outperformed also by the $r$-band magnitude. 

Our interpretation of these trends is as follows: for sources with a single significant peak in the $z$PDF, both $z_{err}$ and $odds$ depend on the width of the peak and produce similar ranks. However, in sources with multiple peaks in the $z$PDF $z_{err}$ underestimates the actual uncertainty in $z_{\rm{best}}$ because it is blind to all but the highest peak. On the other hand, the $odds$ is affected (decreased) by those secondary peaks. The consequence is that the galaxies with multiple peaks in the $z$PDF (which are more likely to have catastrophic redshift errors) get ranked higher by $z_{err}$ compared to $odds$. This has almost no impact on \snmad{} because it is insensitive to outliers, but it shows up in $\eta$, as many more outliers get ranks above the threshold in $z_{err}$ compared to $odds$. 

\subsection{Validation of $z_{err}$ estimates\label{sec:validation-zerr}}

\begin{figure} 
\begin{center}
\includegraphics[width=8.4cm]{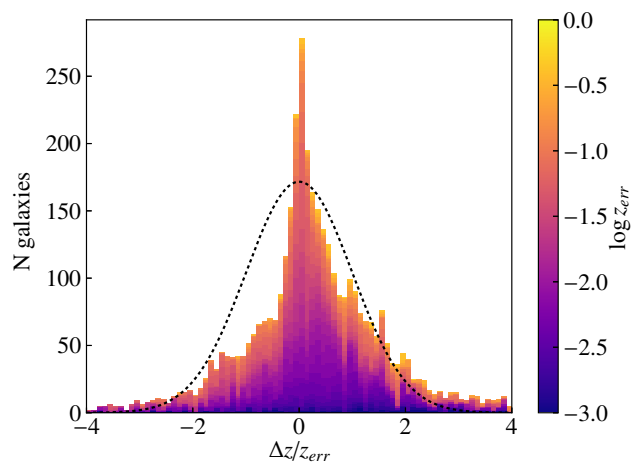}
\end{center}
\caption[]{Distribution of the ratio between the actual redshift error, $\Delta z$, and the uncertainty predicted with the $\Delta\chi^2$ method, $z_{err}$, for individual galaxies in the spectroscopic sample with colour coding for $z_{err}$.\label{fig:reldz-hist}}
\end{figure}

We have shown that $z_{err}$ and $odds$ are the quantities with the stronger correlation to $\Delta z$, and therefore the best choices for the selection of subsamples with the most accurate photo-$z$. In this section, we test if $z_{err}$ corresponds to the actual 1-$\sigma$ uncertainty in $z_{\rm{best}}$ for individual galaxies. 
 
Figure \ref{fig:reldz-hist} shows the distribution of $\Delta z$/$z_{err}$, which corresponds to the error in $z_{\rm{best}}$ in units of the predicted 1-$\sigma$ uncertainty for each galaxy. Under ideal circumstances, the expected distribution is a Gaussian with a standard deviation of 1 (black line). However, the actual distribution is far from Gaussian, with a pointy core and extended tails. 55\% of galaxies are within the [-1,1] interval and 77\% in [-2,2] compared to expectations of 68\% and 95\%, respectively. 
 The colour coding reveals that the core of the distribution is dominated by galaxies with high $z_{err}$ values, while those with small $z_{err}$ dominate in the tails. This suggests that sources with large (small) $z_{err}$ overestimate (underestimate) the actual uncertainty in $z_{\rm{best}}$. 

\begin{figure} 
\begin{center}
\includegraphics[width=8.4cm]{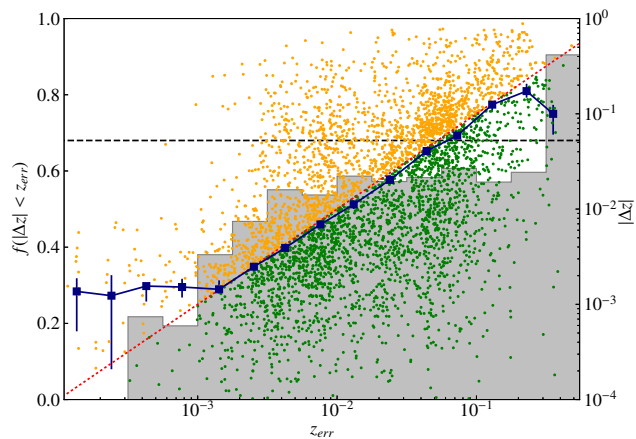}
\end{center}
\caption[]{Distribution of the fraction of galaxies with $\vert\Delta z\vert$ $<$ $z_{err}$ as a function of $z_{err}$ (grey histogram). The horizontal line marks a fraction of 68\% expected if $z_{err}$ estimates are accurate. Small dots indicate the $z_{err}$ and $\vert\Delta z\vert$ of individual galaxies (right hand scale). The dots are coloured orange and green for $\vert\Delta z\vert$ $>$ $z_{err}$ and $\vert\Delta z\vert$ $<$ $z_{err}$, respectively. The dotted red line marks the expected \snmad{} as a function of $z_{err}$, while the connected dark blue squares with error bars indicate the \snmad{} observed and the 16--84$^{th}$ percentiles of its confidence interval.\label{fig:absdz-zerr}}
\end{figure}

Further evidence is presented in Fig. \ref{fig:absdz-zerr} which shows the fraction of galaxies with $\Delta z$$<$$z_{err}$ as a function of $z_{err}$ (grey histogram). If $z_{err}$ estimates were accurate, this fraction should be constant around $\sim$0.68 independently on $z_{err}$. Instead, we find a strong dependence with $z_{err}$ for small $z_{err}$ values with a fraction much lower than the expected 68\%, while the relation becomes flat (but still below the expectation, with $f$$\sim$55--60\%) for $z_{err}$$\gtrsim$0.003. 

One reason for the strong underestimation of the actual redshift uncertainty in sources with low $z_{err}$ is the assumption implicit in the $\Delta\chi^2$ method that the probability distribution for the minimum of $\chi^2$($z$) is the $\chi^2$ distribution for $n$ degrees of freedom \citep{Press92}. For bright sources with high S/N photometry, this is not the case since differences between the intrinsic SED of the galaxy and the closest template in the library are much larger than the photometric errors, implying large values for the reduced $\chi^2$, $\chi_r^2$ = $\chi^2$/($n$-1) $\gg$ 1 \citep[see][for a discussion]{Hernan-Caballero12,Hernan-Caballero15}. 
By contrast, galaxies with low S/N photometry can easily get $\chi_r^2$ $\lesssim$ 1 with multiple redshift-template combinations due to degeneracy in the colour space. Since the model is not linear in the fitting parameters (namely the redshift and spectral type), the $\chi^2$ distribution does not provide a realistic description of the actual redshift uncertainty \citep{Oyaizu08}. 

Another factor contributing to the general overconfidence in $z_{err}$ is the lack of sensitivity in the $\Delta\chi^2$ method to secondary peaks in the distribution of $\chi^2(z)$, and thus, to the probability of a catastrophic redshift error. This is evidenced by the close to 1:1 relation between \snmad{} and $z_{err}$ in the range 0.001--0.1 (blue squares), which shows that for $z_{err}$$>$0.001, it is in fact a realistic prediction of the error in $z_{\rm{best}}$ if catastrophic errors are excluded.

\subsection{Validation of $odds$ estimates\label{sec:validation-odds}}

\begin{figure} 
\begin{center}
\includegraphics[width=8.4cm]{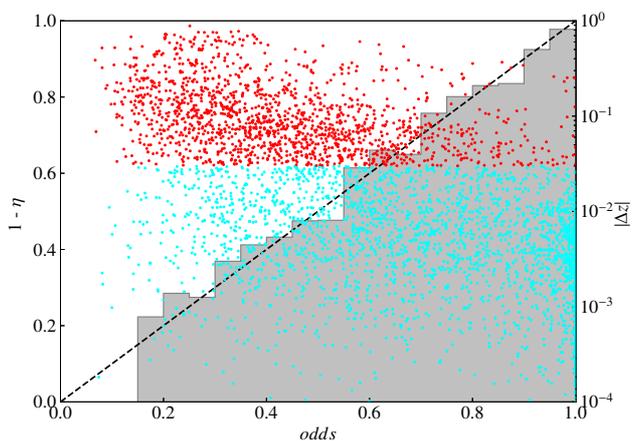}
\end{center}
\caption[]{Distribution of the fraction of sources with $\vert\Delta z\vert$ $<$ 0.03 as a function of $odds$ (grey histogram). The diagonal line marks the 1:1 relation expected if $odds$ estimates for individual galaxies are accurate. Small dots indicate the $odds$ and $\vert\Delta z\vert$ of individual galaxies (right hand scale). The dots are coloured red for outliers ($\vert \Delta z \vert$$>$0.03) or cyan otherwise.\label{fig:odds-zerr}}
\end{figure}

Our definition of the $odds$ parameter (see Sect. \ref{sec:scalar-parameters}) implies that for an individual galaxy, the probability of an error $\vert\Delta z\vert$$>$0.03 in $z_{\rm{best}}$ is $P$ = 1 - $odds$. Therefore, for a sufficiently large subsample, the outlier rate should be $\eta$ $\approx$ 1 - $\langle odds \rangle$.

Since the $z$PDF used to compute the $odds$ is derived from $\chi^2$($z$) and the redshift prior, it is affected by the same $\chi^2_r$$\gg$1 issue that we discussed for $z_{err}$. However, we compensated for this with the contrast correction applied to the $z$PDF in Sect. \ref{sec:contrast-correction}. Also, unlike $z_{err}$, the $odds$ is sensitive to the presence of secondary peaks in the $z$PDF, meaning it should accurately estimate the probability of being an outlier for individual galaxies.
We test this in Fig. \ref{fig:odds-zerr} which shows that the fraction of galaxies with $\vert \Delta z \vert$$<$0.03 has the expected dependence with the $odds$ in the whole range (note that $f$($\vert \Delta z \vert$$<$0.03) = 1 - $\eta$).

\begin{figure} 
\begin{center}
\includegraphics[width=8.4cm]{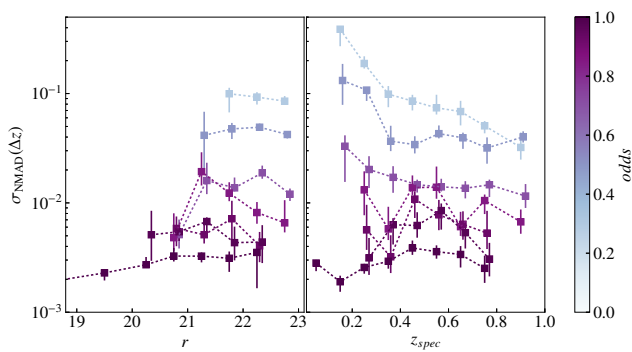}
\end{center}
\caption[]{Dependence of $\sigma_{\rm{NMAD}}$ on the $r$-band magnitude (left) and $z_{\rm{spec}}$ (right) for galaxies grouped in bins of $odds$. $\sigma_{\rm{NMAD}}$ is computed in steps of 0.5 (1 for $r$$<$20) for the $r$-band magnitude and 0.1 for redshift. Only bins containing more than 15 sources are shown.\label{fig:snmad-binsx2-odds}}
\end{figure}

To check whether the magnitude or redshift of the galaxies has any impact on photo-$z$ accuracy that is not already accounted for by the $odds$, we show in Fig. \ref{fig:snmad-binsx2-odds} the magnitude and redshift dependence of $\sigma_{\rm{NMAD}}$ at constant $odds$. 
There is no clear residual dependence of \snmad{} with the $r$-magnitude. However, there seems to be an increase in \snmad{} at low $z$, in particular for sources with very low $odds$. This might be a consequence of the redshift prior favouring intermediate $z_{\rm{best}}$ values in faint sources with low S/N (the prior probability peaks at $z$$\sim$0.6 for $r$=22 and $z$$\sim$0.75 for $r$=23, see Fig. \ref{fig:priors}). Since the number of galaxies with low $z$ and low $odds$ is small, we do not expect this to have a significant impact on our results.

\subsection{Dependence on the spectral type\label{sec:spectral-type}}

Photo-$z$ accuracy is also expected to depend on the spectral type of the galaxies, which determines the contrast of the spectral features that anchor the photo-$z$. The most important such features at the redshifts typical of miniJPAS galaxies are the 4000 \AA{} break of the stellar continuum and the main optical emission lines (H$\alpha$, H$\beta$, [O {\sc ii}] 3727 \AA{} and [O {\sc iii}] 4959+5007 \AA).

The combination of deep broad-band and shallower narrow-band photometry produces some interesting trends for miniJPAS galaxies: at relatively bright magnitudes the narrow-band filters easily detect emission lines, if they are present, which increases the chances of a highly accurate photo-$z$ in star-forming galaxies compared to quiescent ones. However, as we move to fainter magnitudes, the emission lines become increasingly hard to detect, removing the advantage for star-forming galaxies. At magnitudes fainter than the detection limit in the narrow bands, it is quiescent galaxies that often have an edge due to their stronger 4000 \AA{} break which is easily detected in the broad-band photometry.

To quantify the impact of the spectral type on \snmad, we have classified all the galaxies in the spectroscopic sample into two broad categories ``red'' and ``blue'', loosely corresponding to quiescent and star-forming, respectively. 
We repeat the classification three times, according to the value of three different parameters from the best-fitting {\sc cigale} model: (1) $D_n$(4000), which measures the strength of the 4000 \AA{} break using the definition of \citet{Balogh99} and is a proxy for the light-weighted age of the stellar population \citep[e.g.][]{Kauffmann03a,Kauffmann03b,Kriek06,Hernan-Caballero13}; (2) the specific instantaneous star formation rate (sSFR) derived directly from the model star formation history; (3) the equivalent width of the $H\alpha$ line (EW(H$\alpha$), an observational proxy for the sSFR).

Since the boundary between the quiescent and star-forming classes is somewhat arbitrary (there are many galaxies with intermediate properties) and is also redshift- and luminosity-dependent, we simplify the classification by splitting the sample into two same-sized subsamples. The first (second) subsample contains the 50\% of galaxies with the highest (lowest) value of $D_n$(4000), sSFR, or EW(H$\alpha$).
Such criteria do not provide high purity in the resulting samples of quiescent and star-forming galaxies but allow to compare more easily the results between the different selection criteria, and is sufficient to check whether the $odds$ parameter accounts for the dependence on the spectral type. 

\begin{figure} 
\begin{center}
\includegraphics[width=8.4cm]{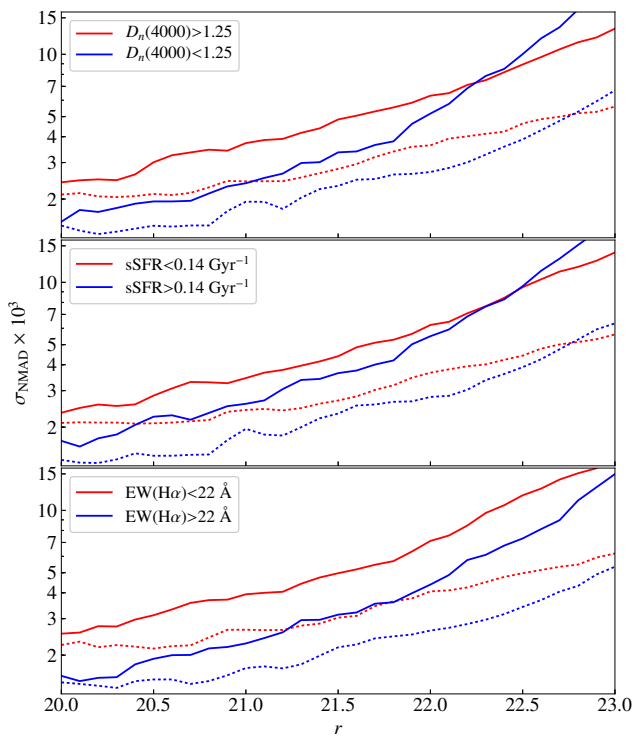}
\end{center}
\caption[]{Variation of $\sigma_{\rm{NMAD}}$ as a function of the cut applied in $r$-band magnitude separately for ``red'' and ``blue''
galaxies. Each panels shows results for a different way to classify the galaxies, using the strength of the 4000 \AA{} break (top), the specific instantaneous SFR (middle), and the equivalent width of H$\alpha$ (bottom). Solid lines correspond to $\sigma_{\rm{NMAD}}$ values at 100\% completeness (no cut in $odds$), while dotted lines indicate $\sigma_{\rm{NMAD}}$ values for a cut in $odds$ corresponding to 50\% completeness.\label{fig:snmad-mag-gtype}}
\end{figure}

Figure \ref{fig:snmad-mag-gtype} shows the \snmad{} for all sources brighter than a given $r$-band magnitude cut in the ``red'' and ``blue'' samples.
The trends are similar irrespective of the quantity used to split the sample but there are some interesting differences: for a cut at very faint magnitudes ($r$$\lesssim$23), the red sample gets lower \snmad{} when the selection is done with $D_n$(4000) or sSFR but not with EW(H$\alpha$). This indicates that a substantial number of galaxies switches between the red and blue classes depending on the parameter used. Classifying with EW(H$\alpha$) puts some high $D_n$(4000) galaxies (that for some reason, maybe active nuclei,  also have H$\alpha$ emission) in the blue sample, improving its \snmad. 

A consequence of this is that the photo-$z$ accuracy at a given magnitude depends not only on the spectral type but on how the spectral type is defined. In particular, more restrictive classification criteria are likely to improve the photo-$z$ accuracy for both quiescent and star-forming galaxies, since it is intermediate cases (where neither the 4000 \AA{} break nor emission lines are strong) that constitute the most difficult targets for photo-$z$ calculation.

\begin{figure} 
\begin{center}
\includegraphics[width=8.4cm]{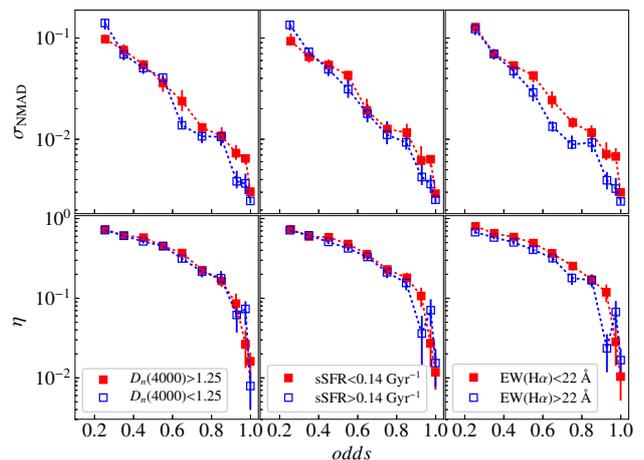}
\end{center}
\caption[]{Dependence of $\sigma_{\rm{NMAD}}$ and $\eta$ with $odds$ for ``quiescent'' (red solid symbols) and ``star-forming'' (blue open symbols) galaxies in the spectroscopic sample according to three different classification criteria (left, center, and right panels, see text for details).  The error bars represent 1-$\sigma$ confidence intervals calculated with bootstrap resampling.\label{fig:snmad-outrate-odds-gtype}}
\end{figure}

The factors responsible for the different photo-$z$ accuracy in quiescent and star-forming galaxies are also reflected in the value of the $odds$ parameter. Figure \ref{fig:snmad-outrate-odds-gtype} shows that if the comparison is made between galaxies within a narrow range of $odds$, the difference in \snmad{} between red and blue galaxies disappears. The similarity is even stronger for $\eta$, indicating that the $odds$ parameter accurately predicts the probability of a catastrophic error for both quiescent and star-forming galaxies. 

\subsection{Representativeness of the spectroscopic sample}

The photo-$z$ performance statistics presented so far refer only to the subsample of miniJPAS galaxies with spectroscopic redshifts ($s$-sample). We can expect these statistics to also predict the performance in the sources with no spectroscopic redshifts ($p$-sample) only if the $s$- and $p$-sample are drawn from the same parent population.

\begin{figure} 
\begin{center}
\includegraphics[width=8.4cm]{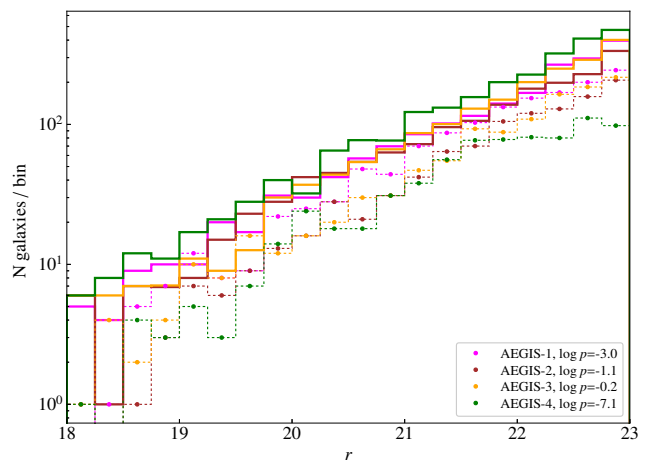}
\end{center}
\caption[]{Distribution of the observed $r$-band magnitude of miniJPAS galaxies for the photo-$z$-only sample (solid lines) and the spectroscopic sample (connected dots) for individual pointings. $\log$ $p$-values in the legend indicate the probability of the spectroscopic and photo-$z$-only samples being drawn from the same parent population.\label{fig:mag-distrib-tiles}}
\end{figure}

We already showed in Fig. \ref{fig:Nmag} that the distribution of $r$-band magnitude in the $s$-sample reproduces that of the whole miniJPAS when every source is weighted with its probability of being a galaxy, $P_G$.
Figure \ref{fig:mag-distrib-tiles} presents a more detailed comparison of the magnitude distribution in the $s$-sample and $p$-sample. Since the spectroscopic coverage and the depth of the miniJPAS images varies from one pointing to another, we show the distributions for each pointing separately. We use the two-sample Kolmogorov-Smirnov test to calculate the probability $p$ of the $s$-sample and $p$-sample being drawn from the same parent population. The difference between samples is significant ($p$$<$0.01) only for pointing AEGIS-4. Unlike the other pointings, the spectroscopic counts in AEGIS-4 flatten at $r$$>$21.5. Interestingly, this pointing has the smallest area covered by DEEP and the lowest number of spectroscopic galaxies and, as a consequence, also the highest counts of photo-$z$-only galaxies in almost every magnitude bin. The presence of a large galaxy cluster centred at RA=213.6254, DEC=51.9379 (see Fig. 28 in \citetalias{Bonoli21}) may also boost the galaxy counts in AEGIS-4. 

\begin{figure} 
\begin{center}
\includegraphics[width=8.4cm]{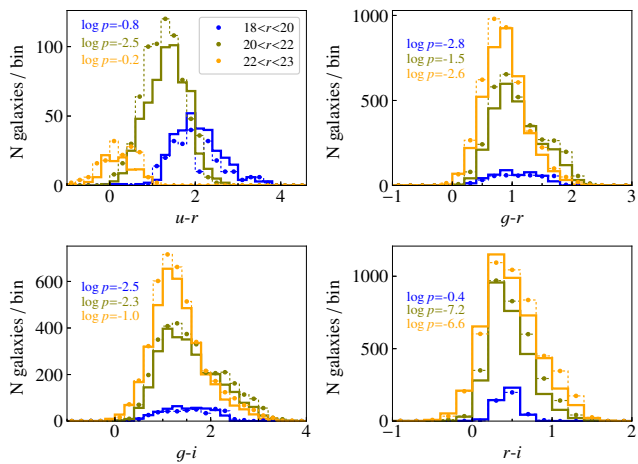}
\end{center}
\caption[]{Distribution of the observed broad-band colours in three bins of magnitude. Each plot includes only sources with $>$3$\sigma$ detections in the two bands defining the colour index and with valid photo-$z$ estimates. Solid lines show the distributions measured for galaxies in the photo-$z$-only sample. Connected dots represent the distribution for galaxies in the spectroscopic sample (scaled by a factor 2 to facilitate the comparison).\label{fig:color-distrib}}
\end{figure}

In Fig. \ref{fig:color-distrib}, we compare the distributions of four broad-band colours for galaxies in the $p$-sample and $s$-sample. For each colour index, only sources with $>$3$\sigma$ detection in both bands are considered.
We show the distributions separately for three magnitude ranges. 
The magnitude-dependence of the colour distribution is small except for the $u$-$r$ colour, where the S/N$>$3 requirement for the $u$ band implies that most red objects are not selected at faint magnitudes, shifting the distribution towards bluer $u$-$r$.
Discrepancies between distributions for the $s$-sample and $p$-sample are highly significant, particularly for the $r$-$i$ colour in $r$$>$20 sources ($p$$<$0.0001). However, the range of colour indices covered are the same and, in all cases, the mean colour index of the distribution differs by less than 0.1 magnitudes (Table \ref{table:average-color}).

\begin{table}[] 
\centering
\caption{Average colour indices by magnitude\label{table:average-color}} 
\begin{tabular}{ccccccc} 
\hline
\hline
\multicolumn{1}{c}{colour} & \multicolumn{2}{c}{18$<$$r$$<$20} & \multicolumn{2}{c}{20$<$$r$$<$22} & \multicolumn{2}{c}{22$<$$r$$<$23}\\
\multicolumn{1}{c}{index} & \multicolumn{1}{c}{phot} & \multicolumn{1}{c}{spec} & \multicolumn{1}{c}{phot} & \multicolumn{1}{c}{spec} & \multicolumn{1}{c}{phot} & \multicolumn{1}{c}{spec} \\
\hline       
$\langle$$u$-$r$$\rangle$ & 2.051 & 2.028 &  1.351 & 1.257 &  0.208 & 0.201 \\ 
$\langle$$g$-$r$$\rangle$ & 1.035 & 1.123 &  1.086 & 1.096 &  0.875 & 0.871 \\ 
$\langle$$g$-$i$$\rangle$ & 1.479 & 1.578 &  1.552 & 1.590 &  1.307 & 1.345 \\ 
$\langle$$r$-$i$$\rangle$ & 0.444 & 0.455 &  0.469 & 0.498 &  0.480 & 0.520 \\ 
\hline
\end{tabular}
\end{table}

\begin{figure} 
\begin{center}
\includegraphics[width=8.4cm]{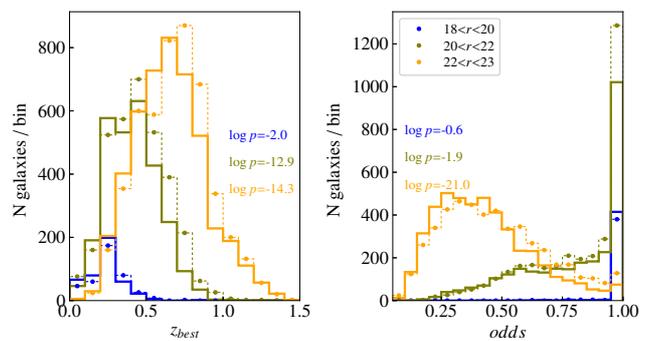}
\end{center}
\caption[]{Distribution of $z_{\rm{best}}$ and $odds$ in bins of magnitude for the photo-$z$-only and spectroscopic samples. Symbols as in Fig. \ref{fig:color-distrib}.\label{fig:zbest-odds-distrib-mag}}
\end{figure}

The distribution of $z_{\rm{best}}$ (left panel in Fig. \ref{fig:zbest-odds-distrib-mag}) is slightly biased towards higher values for $r$$>$20 sources in the $s$-sample compared to the $p$-sample. This is consistent with their redder $g$-$i$ and $r$-$i$ colours. For the $odds$ parameter (right panel), the difference is significant only among the faintest galaxies ($r$$>$22) and indicates that $r$$>$22 galaxies have higher $odds$, on average, in the $s$-sample compared to the $p$-sample.
This may be a consequence of the requirement of a high confidence in the spectroscopic redshift for selection into the $s$-sample. At faint magnitudes, such confidence requires high contrast spectral features (a strong 4000 \AA{} break or emission lines) that also help increase the confidence in the photo-$z$ estimate. 

As a result of this analysis, we conclude that the $s$-sample is slightly biased towards redder and more distant galaxies but also with higher $odds$ compared to same-magnitude galaxies in the $p$-sample. However, such small differences do not suggest the existence of a population of galaxies in the $p$-sample that is not represented in the $s$-sample. Accordingly, we consider the templates selected for the $s$-sample to also be suitable for the $p$-sample and we expect a similar performance.

The small but highly significant differences in the $odds$ distributions are important since the strong correlation between $odds$ and photo-$z$ accuracy implies that for magnitude-limited subsamples the \snmad{} and $\eta$ measured in the $s$-sample probably overestimate the actual accuracy in the $p$-sample. 
However, we have shown that $odds$ is the best predictor of photo-$z$ accuracy and that, at constant $odds$, the residual dependence of photo-$z$ accuracy with $r$, $z_{\rm{spec}}$, or the spectral type is small. 

This implies that the distribution of $\vert\Delta z\vert$ at a given $odds$ obtained for the $s$-sample should also represent that of the $p$-sample. In the next section we show how to use this to compensate for the different $odds$ distributions in the $s$- and $p$-sample to obtain realistic photo-$z$ performance statistics for the $p$-sample.

\subsection{Extrapolation to the entire miniJPAS sample}

\begin{figure} 
\begin{center}
\includegraphics[width=8.4cm]{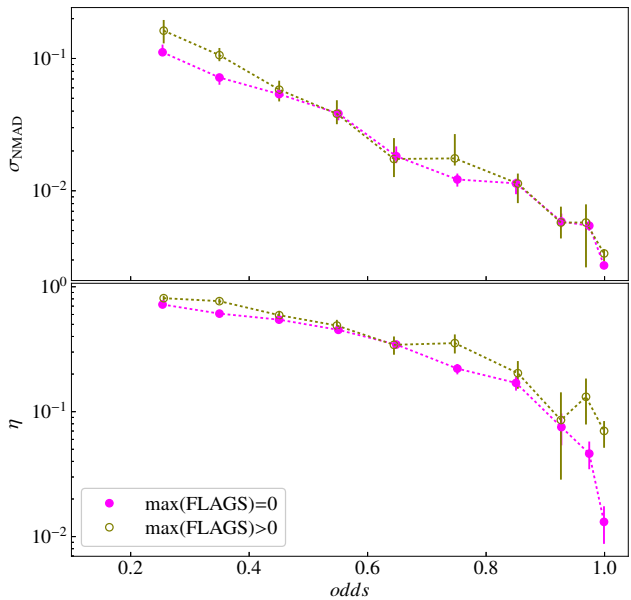}
\end{center}
\caption[]{Dependence of \snmad{} (top) and $\eta$ (bottom) with $odds$ for sources in the spectroscopic sample with no photometry flags in any bands (solid symbols) and sources with flags in one or more bands (open symbols).\label{fig:snmad-flags}}
\end{figure}

The spectroscopic sample used for all the analysis up to this point excludes sources with flags in the photometry. 
In order to check the impact of photometric flags in the photo-$z$ accuracy, we compare in Fig. \ref{fig:snmad-flags} the values of \snmad{} and $\eta$ as a function of $odds$ for flagged and non-flagged galaxies in the spectroscopic sample. The differences in \snmad{} are minimal and largely consistent within the statistical errors except at very low $odds$ ($odds$$<$0.4). This suggest that most flagged sources are barely affected in their photo-$z$ accuracy by the photometric issue signalled by the flags. 
The trends for $\eta$ are also very similar except at $odds$$>$0.9, where the outlier rate stays at $\eta$$\sim$0.1 for flagged sources while it falls down to $\eta$$\sim$0.01 for non-flagged ones. This clear excess of outliers at very high $odds$ is probably a consequence of strong spurious peaks in the $z$PDF caused by artefacts in the photometry of some of the flagged sources. 
Since flagged sources represent $\sim$30\% of the miniJPAS sample, we consider that such a small increase in the outlier rate relative to non-flagged ones does not justify purging flagged sources unless very high reliability (very low $\eta$) is required.

The strong dependence of the distribution of $\vert\Delta z\vert$ with $odds$ (Fig. \ref{fig:absdz}) and the lack of a significant residual dependence with other quantities at constant $odds$ allow us to estimate the number of galaxies with photo-$z$ errors below a given threshold in any arbitrary sample of miniJPAS sources: 

\begin{equation}\label{eqn:numbers}
N(\vert\Delta z\vert<\Delta z') = \sum\limits_i P_{G,i} ~ f_S(\vert\Delta z\vert<\Delta z'~\vert ~odds_i)
\end{equation}
\noindent where $P_{G,i}$ is the probability of being a galaxy for source $i$ (see Sect. \ref{sec:star-galaxy-class}) and
$f_S$($\vert\Delta z\vert$$<$$\Delta z'$\,$\vert$\,$odds_i$) is the frequency of $\vert\Delta z\vert$$<$$\Delta z'$ among galaxies with $odds$$\approx$$odds_i$ in the spectroscopic sample. 
Table \ref{table:dzX-odds} lists the values of $f_S$($\vert\Delta z\vert<\Delta z'$ $\vert$ $odds$) calculated for intervals of $odds$ with several values of the threshold $\Delta z'$.

\begin{figure} 
\begin{center}
\includegraphics[width=8.4cm]{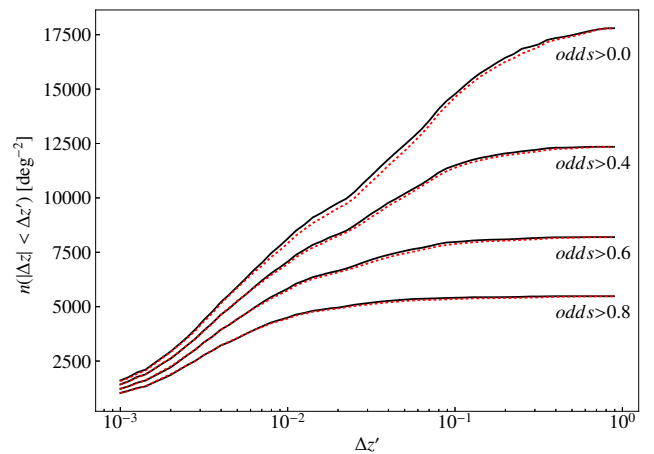}
\end{center}
\caption[]{Predicted density of $r$$<$23 miniJPAS galaxies with redshift errors $\vert \Delta z \vert$$<$$\Delta z'$ as a function of the threshold $\Delta z'$ for four different cuts in the $odds$ parameter. The solid lines are generated by applying the same value of $f_S$($\vert\Delta z\vert$$<$$\Delta z'$\,$\vert$\,$odds_i$) calculated in non-flagged sources to both flagged and non-flagged ones, while the dotted lines uses separate values for flagged and non-flagged sources.\label{fig:predict-Ndz}}
\end{figure}

We use Eq. \ref{eqn:numbers} to compute the number density $n$ of galaxies in miniJPAS with $r$$<$23 and $\vert\Delta z\vert$$<$$\Delta z'$ for several cuts in $odds$ (Fig. \ref{fig:predict-Ndz}). For this we assume that the effective area of miniJPAS (after taking the masked areas into account) is 0.895 deg$^2$ (see \citetalias{Bonoli21}). We repeat the calculation twice: first using the frequencies $f_S$($\vert\Delta z\vert<\Delta z'$ $\vert$ $odds$) calculated on non-flagged sources for flagged and non-flagged ones indistinctly; second using their own $f_S$($\vert\Delta z\vert<\Delta z'$ $\vert$ $odds$) for both flagged and non-flagged sources.
The difference in number counts between the two methods is $<$3\% for any $\Delta z'$ and cut in $odds$.

These distributions show that there are $\sim$17,500 galaxies per deg$^2$ in miniJPAS at $r$$<$23, of which $\sim$4,200 have $\vert\Delta z\vert$$<$0.003. However, selecting all of them requires to put no constraint in $odds$, which results in large average errors (\snmad{} = 0.013) and high rate of outliers ($\eta$=0.39). The targeted photo-$z$ accuracy for J-PAS \citep[\snmad=0.003;][]{Benitez14} is achieved after imposing $odds$$>$0.82, that implies selecting $\sim$5,200 galaxies per deg$^2$ (of which $\sim$2,600 have $\vert\Delta z\vert$$<$0.003 and only $\sim$5\% are outliers).

\begin{table}[] 
\centering
\caption{Fraction of sources with $\vert\Delta z\vert$$<\Delta z'$ per $odds$ interval in the spectroscopic sample} 
\resizebox{\columnwidth}{!}{ 
\begin{tabular}{crcccc} 
\hline
\hline
\multirow{1}{*}{$odds$} & 
\multirow{1}{*}{$N_{spec}$} & 
\multirow{1}{*}{$\Delta z'$=0.003} &
\multirow{1}{*}{$\Delta z'$=0.01} & 
\multirow{1}{*}{$\Delta z'$=0.03} &
\multirow{1}{*}{$\Delta z'$=0.1}\\
\hline
0.025--0.075  &   3 &  0.333 & 0.333 & 0.333 & 0.667\\
0.075--0.125  &  29 &  0.103 & 0.172 & 0.207 & 0.379\\
0.125--0.175  & 102 &  0.059 & 0.157 & 0.216 & 0.441\\
0.175--0.225  & 150 &  0.080 & 0.160 & 0.270 & 0.533\\
0.225--0.275  & 210 &  0.055 & 0.133 & 0.227 & 0.510\\
0.275--0.325  & 244 &  0.074 & 0.213 & 0.369 & 0.631\\
0.325--0.375  & 249 &  0.121 & 0.265 & 0.402 & 0.735\\
0.375--0.425  & 249 &  0.113 & 0.221 & 0.378 & 0.763\\
0.425--0.475  & 247 &  0.115 & 0.296 & 0.489 & 0.838\\
0.475--0.525  & 250 &  0.109 & 0.284 & 0.475 & 0.852\\
0.525--0.575  & 239 &  0.159 & 0.326 & 0.524 & 0.879\\
0.575--0.625  & 235 &  0.151 & 0.396 & 0.641 & 0.906\\
0.625--0.675  & 205 &  0.210 & 0.454 & 0.650 & 0.941\\
0.675--0.725  & 198 &  0.180 & 0.485 & 0.703 & 0.939\\
0.725--0.775  & 145 &  0.239 & 0.572 & 0.801 & 0.966\\
0.775--0.825  & 189 &  0.236 & 0.550 & 0.805 & 0.963\\
0.825--0.875  & 159 &  0.259 & 0.566 & 0.827 & 0.962\\
0.875--0.925  & 155 &  0.378 & 0.710 & 0.882 & 0.981\\
0.925--0.975  & 223 &  0.355 & 0.767 & 0.956 & 0.996\\
0.975--1.000  & 827 &  0.596 & 0.921 & 0.981 & 0.993\\
\hline
\end{tabular}}
\label{table:dzX-odds}
\end{table}

\section{Summary}

This paper describes the procedures that we followed in order to generate the photo-$z$ catalogue of miniJPAS, a $\sim$1 deg$^2$ imaging survey in 60 optical bands encompassing the AEGIS field. We also provide a detailed analysis of the photo-$z$ performance enabled by the exhaustive spectroscopic coverage of AEGIS by the SDSS and DEEP surveys.

We rely on forced (dual mode) photometry obtained for a $r$-band selected catalogue with matched apertures corresponding to 1 Kron radius in the $r$ band (restricted AUTO aperture), with corrections to compensate for the difference in the PSF of each band with respect to $r$ (PSFCOR photometry).

We use {\sc {\sc cigale}} to generate stellar population synthesis models matching the photo-spectra of all $r$$<$22 galaxies with spectroscopic redshifts. Synthetic photometry obtained from these models is used to identify systematic offsets in the observed photometry with respect to the models. We show that an iterative correction of these offsets converges to the same solution within $\sim$0.01 magnitudes irrespective of the initial offsets.

The photo-$z$ are computed with a customised version of {\sc LePhare}. The spectral templates used are the best-fitting {\sc {\sc cigale}} models of 50 miniJPAS galaxies, selected from a larger set of 455 candidates as the ones that produce the most accurate photo-$z$ in a test sample. 

We show that the redshift probability distribution functions ($z$PDF) generated by {\sc LePhare} are slightly overconfident for $r$$<$19 galaxies but severely under-confident at $r$$>$20. A simple contrast correction of the $z$PDF compensates for this magnitude dependence. 

Comparison between the mode of the $z$PDF ($z_{\rm{best}}$) and the spectroscopic redshift ($z_{\rm{spec}}$) shows that the distribution of absolute redshift errors ($\vert\Delta z\vert$) is bimodal. The main peak at $\vert\Delta z\vert$$\sim$0.004 corresponds to the typical inaccuracy in photo-$z$ estimates while a second peak at $\vert\Delta z\vert$$\sim$0.04 represents catastrophic errors.

The distribution of $z_{\rm{best}}$ for the galaxies in miniJPAS closely follows the distribution of $z_{\rm{spec}}$ in the spectroscopic subsample. This indicates that (1) the spectroscopic sample is representative of the redshift distribution of the whole sample and (2) we successfully detect variation in the density of galaxies along the redshift dimension, at least up to $z$$\sim$0.8.

We find $z_{\rm{best}}$ estimates are biased towards $z_{\rm{best}}$$>$$z_{\rm{spec}}$ by $\sim$0.10--0.14\%. This bias is consistent among the four miniJPAS pointings and shows no clear dependence with either the magnitude or the redshift of the sources. In spite of our efforts to identify the origin of this bias, it remains uncertain and demands further analysis.

The $odds$ parameter has the strongest correlation to $\vert\Delta z\vert$ of any photo-$z$ related quantities. In particular, its correlation is significantly stronger than obtained with the predicted 1-$\sigma$ redshift error ($z_{err}$) since the latter is insensitive to the probability of a catastrophic redshift error. This implies that applying a cut in $odds$ is the most efficient way to select a fixed-size subsample with the best possible photo-$z$ (or, conversely, the largest sample within some photo-$z$ quality constraints).
We also show that there is no clear residual dependence of the photo-$z$ accuracy with the $r$-band magnitude at constant $odds$, while there seems to be a redshift dependence for sources with very low $odds$, consistent with the expected effect of the redshift prior in low S/N photometry.

Comparison of $z_{err}$ and $\vert\Delta z\vert$ shows that the former increasingly underestimates the actual errors in $z_{\rm{best}}$ for lower $z_{err}$ values. On the other hand, the $odds$ parameter accurately represents the probability of a redshift outlier ($\vert\Delta z\vert$$>$0.03).

The photo-$z$ accuracy is dependent on the spectral type. Emission lines allow star-forming galaxies to obtain lower \snmad{} compared to quiescent ones at bright magnitudes but their advantage vanishes at faint magnitudes as the emission lines become increasingly hard to detect and the photo-$z$ solution becomes dominated by the 4000 \AA{} break. The dependence on the spectral type disappears if \snmad{} or $\eta$ are calculated for sources within narrow intervals of $odds$.

We confirm that the distribution of magnitudes and broadband colours in the spectroscopic sample is roughly consistent with the photo-$z$-only sample, albeit galaxies in the latter are slightly bluer ($\lesssim$0.1 magnitudes) and have slightly lower $z_{\rm{best}}$ on average. The distribution of $odds$ for $r$$>$22 galaxies is also biased towards higher values in the spectroscopic sample. We take this into account to generate realistic estimates of the expected photo-$z$ accuracy in the photo-$z$-only sample. 

We conclude that at the depth of miniJPAS, there are $\sim$17,500 galaxies per deg$^2$ with valid photo-$z$ estimates, of which $\sim$4,200 have $\vert\Delta z\vert$$<$0.003. The typical error for $r$$<$23 galaxies is \snmad=0.013 with an outlier rate $\eta$=0.39. The target photo-$z$ accuracy  \snmad=0.003 is achieved after imposing $odds$$>$0.82. Under such constraint, the density of galaxies selected is reduced by 70\% to $n$$\sim$5,200 deg$^{-2}$ (of which $\sim$2,600 have $\vert\Delta z\vert$$<$0.003) and the outlier rate decreases to $\eta$=0.05.

\begin{acknowledgements}
We thank the anonymous referee for useful comments and suggestions that helped improve this work.
This paper has gone through internal review by the J-PAS collaboration.
Based on observations made with the JST/T250 telescope at the Observatorio Astrof\'isico de Javalambre (OAJ), in Teruel, owned, managed, and operated by the Centro de Estudios de F\'isica del Cosmos de Arag\'on (CEFCA). We acknowledge the OAJ Data Processing
and Archiving Unit (UPAD) for reducing and calibrating the OAJ data used in this work.
Funding for the J-PAS Project has been provided by the Governments of Spain and Arag\'on through the Fondo de Inversi\'on de Teruel, European FEDER funding and the Spanish Ministry of Science, Innovation and Universities, and by the Brazilian agencies FINEP, FAPESP, FAPERJ and by the National Observatory of Brazil. Additional funding was also provided by the Tartu Observatory and by the J-PAS Chinese Astronomical Consortium.
Funding for OAJ, UPAD, and CEFCA has been provided by the Governments of Spain and Arag\'on through the Fondo de Inversiones de Teruel; the Arag\'on Government through the Research Groups E96, E103, and E16\_17R; the Spanish Ministry of Science, Innovation and Universities (MCIU/AEI/FEDER, UE) with grant PGC2018-097585-B-C21; the Spanish Ministry of Economy and Competitiveness (MINECO/FEDER, UE) under AYA2015-66211-C2-1-P, AYA2015-66211-C2-2, AYA2012-30789, and ICTS-2009-14; and European FEDER funding (FCDD10-4E-867, FCDD13-4E-2685).
C.Q. acknowledges support from Brazilian agencies FAPESP and CAPES. E.S.C. acknowledges financial support from Brazilian agencies CNPq and FAPESP (process \#2019/19687-2). L.D.G, R.G.D. and G.M.S. acknowledge support from the State Agency for Research of the Spanish MCIU through the "Center of Excellence Severo Ochoa" award to the Instituto de Astrof\'isica de Andaluc\'ia (SEV-2017-0709) and the projects PID2019-109067-GB100 and AYA2016-77846-P.

Part of this work was supported by institutional research funding IUT40-2, JPUT907 and PRG1006 of the Estonian Ministry of Education and Research. We acknowledge the support by the Centre of Excellence ``Dark side of the Universe'' (TK133) financed by the European Union through the European Regional Development Fund.
\end{acknowledgements}

\begin{table*}[] 
\centering
\caption{zeropoint recalibration offsets [mag]} 
\begin{tabular}{crrrr} 
\hline
\hline
\multirow{1}{*}{band} & 
\multirow{1}{*}{AEGIS-1} & 
\multirow{1}{*}{AEGIS-2} &
\multirow{1}{*}{AEGIS-3} & 
\multirow{1}{*}{AEGIS-4} \\
\hline
uJAVA &  0.200 $\pm$  0.043 &  0.099 $\pm$  0.044 &  0.056 $\pm$  0.064 &  0.039 $\pm$  0.051\\ 
J0378 &  0.014 $\pm$  0.042 &  0.015 $\pm$  0.040 &  0.161 $\pm$  0.089 & -0.139 $\pm$  0.075\\ 
J0390 &  0.331 $\pm$  0.023 &  0.114 $\pm$  0.033 &  0.124 $\pm$  0.045 &  0.032 $\pm$  0.043\\ 
J0400 &  0.225 $\pm$  0.028 &  0.090 $\pm$  0.035 &  0.163 $\pm$  0.050 &  0.250 $\pm$  0.041\\ 
J0410 &  0.196 $\pm$  0.035 &  0.119 $\pm$  0.042 &  0.235 $\pm$  0.050 & -0.006 $\pm$  0.057\\ 
J0420 &  0.230 $\pm$  0.037 &  0.040 $\pm$  0.046 &  0.035 $\pm$  0.063 &  0.008 $\pm$  0.058\\ 
J0430 &  0.105 $\pm$  0.034 &  0.099 $\pm$  0.032 &  0.119 $\pm$  0.060 &  0.058 $\pm$  0.030\\ 
J0440 &  0.439 $\pm$  0.029 &  0.156 $\pm$  0.039 &  0.141 $\pm$  0.045 &  0.084 $\pm$  0.038\\ 
J0450 &  0.082 $\pm$  0.039 & -0.017 $\pm$  0.051 &  0.050 $\pm$  0.074 &  0.065 $\pm$  0.051\\ 
J0460 &  0.192 $\pm$  0.015 &  0.160 $\pm$  0.021 &  0.103 $\pm$  0.044 &  0.123 $\pm$  0.030\\ 
J0470 &  0.191 $\pm$  0.021 &  0.109 $\pm$  0.027 &  0.166 $\pm$  0.035 &  0.336 $\pm$  0.034\\ 
J0480 &  0.232 $\pm$  0.022 &  0.131 $\pm$  0.032 &  0.206 $\pm$  0.032 &  0.119 $\pm$  0.032\\ 
J0490 &  0.219 $\pm$  0.026 &  0.072 $\pm$  0.037 &  0.070 $\pm$  0.040 &  0.105 $\pm$  0.057\\ 
J0500 &  0.080 $\pm$  0.015 &  0.098 $\pm$  0.026 &  0.101 $\pm$  0.031 &  0.070 $\pm$  0.024\\ 
J0510 &  0.416 $\pm$  0.023 &  0.156 $\pm$  0.026 &  0.095 $\pm$  0.023 &  0.047 $\pm$  0.025\\ 
J0520 &  0.042 $\pm$  0.031 & -0.042 $\pm$  0.028 &  0.082 $\pm$  0.037 &  0.012 $\pm$  0.035\\ 
J0530 &  0.234 $\pm$  0.016 &  0.065 $\pm$  0.021 &  0.038 $\pm$  0.016 &  0.069 $\pm$  0.038\\ 
J0540 &  0.174 $\pm$  0.019 &  0.038 $\pm$  0.021 &  0.130 $\pm$  0.023 &  0.175 $\pm$  0.025\\ 
J0550 &  0.145 $\pm$  0.019 &  0.085 $\pm$  0.027 &  0.127 $\pm$  0.022 &  0.078 $\pm$  0.031\\ 
J0560 &  0.209 $\pm$  0.018 &  0.024 $\pm$  0.025 &  0.047 $\pm$  0.029 &  0.037 $\pm$  0.041\\ 
J0570 &  0.086 $\pm$  0.016 &  0.051 $\pm$  0.023 &  0.067 $\pm$  0.022 &  0.044 $\pm$  0.028\\ 
J0580 &  0.403 $\pm$  0.017 &  0.070 $\pm$  0.021 &  0.088 $\pm$  0.016 &  0.046 $\pm$  0.021\\ 
J0590 &  0.121 $\pm$  0.026 & -0.024 $\pm$  0.016 &  0.061 $\pm$  0.028 & -0.017 $\pm$  0.027\\ 
J0600 &  0.126 $\pm$  0.017 &  0.065 $\pm$  0.019 &  0.131 $\pm$  0.019 &  0.043 $\pm$  0.026\\ 
J0610 &  0.112 $\pm$  0.017 &  0.063 $\pm$  0.019 &  0.109 $\pm$  0.016 &  0.182 $\pm$  0.019\\ 
J0620 &  0.178 $\pm$  0.021 &  0.084 $\pm$  0.021 &  0.110 $\pm$  0.019 &  0.052 $\pm$  0.031\\ 
J0630 &  0.185 $\pm$  0.015 &  0.015 $\pm$  0.021 &  0.026 $\pm$  0.020 &  0.045 $\pm$  0.031\\ 
J0640 &  0.101 $\pm$  0.016 &  0.054 $\pm$  0.019 &  0.103 $\pm$  0.011 &  0.016 $\pm$  0.020\\ 
J0650 &  0.335 $\pm$  0.017 &  0.087 $\pm$  0.020 &  0.042 $\pm$  0.010 &  0.045 $\pm$  0.019\\ 
J0660 &  0.133 $\pm$  0.010 &  0.102 $\pm$  0.016 &  0.129 $\pm$  0.007 &  0.089 $\pm$  0.031\\ 
J0670 &  0.253 $\pm$  0.015 &  0.056 $\pm$  0.014 &  0.089 $\pm$  0.007 &  0.102 $\pm$  0.024\\ 
J0680 &  0.099 $\pm$  0.016 &  0.074 $\pm$  0.016 &  0.082 $\pm$  0.008 &  0.109 $\pm$  0.033\\ 
J0690 &  0.151 $\pm$  0.018 &  0.076 $\pm$  0.021 &  0.076 $\pm$  0.019 &  0.031 $\pm$  0.032\\ 
J0700 &  0.172 $\pm$  0.012 &  0.028 $\pm$  0.020 &  0.032 $\pm$  0.012 &  0.015 $\pm$  0.021\\ 
J0710 &  0.108 $\pm$  0.017 &  0.015 $\pm$  0.019 &  0.093 $\pm$  0.012 &  0.018 $\pm$  0.027\\ 
J0720 &  0.232 $\pm$  0.022 &  0.103 $\pm$  0.020 &  0.033 $\pm$  0.008 &  0.052 $\pm$  0.019\\ 
J0730 &  0.159 $\pm$  0.019 & -0.076 $\pm$  0.023 &  0.024 $\pm$  0.015 & -0.034 $\pm$  0.022\\ 
J0740 &  0.309 $\pm$  0.017 &  0.022 $\pm$  0.019 &  0.086 $\pm$  0.013 &  0.057 $\pm$  0.036\\ 
J0750 &  0.100 $\pm$  0.016 &  0.062 $\pm$  0.020 &  0.083 $\pm$  0.011 &  0.143 $\pm$  0.012\\ 
J0760 &  0.143 $\pm$  0.018 &  0.125 $\pm$  0.024 &  0.086 $\pm$  0.019 &  0.045 $\pm$  0.038\\ 
J0770 &  0.190 $\pm$  0.019 &  0.045 $\pm$  0.021 &  0.054 $\pm$  0.016 &  0.065 $\pm$  0.026\\ 
J0780 &  0.164 $\pm$  0.017 &  0.056 $\pm$  0.024 &  0.099 $\pm$  0.018 &  0.055 $\pm$  0.034\\ 
J0790 &  0.170 $\pm$  0.014 &  0.102 $\pm$  0.022 & -0.006 $\pm$  0.012 &  0.053 $\pm$  0.021\\ 
J0800 &  0.044 $\pm$  0.016 &  0.039 $\pm$  0.023 &  0.029 $\pm$  0.026 &  0.014 $\pm$  0.013\\ 
J0810 &  0.047 $\pm$  0.015 &  0.027 $\pm$  0.025 &  0.023 $\pm$  0.025 &  0.001 $\pm$  0.012\\ 
J0820 &  0.049 $\pm$  0.021 &  0.007 $\pm$  0.022 &  0.062 $\pm$  0.026 &  0.031 $\pm$  0.019\\ 
J0830 &  0.095 $\pm$  0.018 &  0.068 $\pm$  0.023 &  0.052 $\pm$  0.018 &  0.058 $\pm$  0.023\\ 
J0840 &  0.126 $\pm$  0.022 &  0.098 $\pm$  0.023 &  0.036 $\pm$  0.023 &  0.061 $\pm$  0.020\\ 
J0850 &  0.103 $\pm$  0.023 &  0.057 $\pm$  0.024 &  0.014 $\pm$  0.025 &  0.008 $\pm$  0.020\\ 
J0860 &  0.117 $\pm$  0.017 &  0.050 $\pm$  0.006 &  0.078 $\pm$  0.015 &  0.036 $\pm$  0.035\\ 
J0870 &  0.300 $\pm$  0.022 &  0.067 $\pm$  0.023 &  0.241 $\pm$  0.017 &  0.036 $\pm$  0.037\\ 
J0880 &  0.293 $\pm$  0.022 &  0.177 $\pm$  0.024 &  0.240 $\pm$  0.029 &  0.031 $\pm$  0.047\\ 
J0890 &  0.247 $\pm$  0.021 &  0.182 $\pm$  0.032 &  0.213 $\pm$  0.021 &  0.178 $\pm$  0.041\\ 
J0900 &  0.173 $\pm$  0.018 &  0.221 $\pm$  0.030 &  0.183 $\pm$  0.019 &  0.240 $\pm$  0.040\\ 
J0910 &  0.128 $\pm$  0.020 &  0.297 $\pm$  0.027 &  0.203 $\pm$  0.026 &  0.143 $\pm$  0.038\\ 
J1007 &  0.179 $\pm$  0.018 &  0.182 $\pm$  0.024 &  0.274 $\pm$  0.045 &  0.074 $\pm$  0.037\\ 
uJPAS &  0.111 $\pm$  0.036 &  0.162 $\pm$  0.036 &  0.156 $\pm$  0.058 &  0.002 $\pm$  0.048\\ 
gSDSS &  0.092 $\pm$  0.011 &  0.060 $\pm$  0.016 &  0.154 $\pm$  0.013 &  0.095 $\pm$  0.019\\ 
rSDSS &  0.000 $\pm$  0.000 &  0.000 $\pm$  0.000 &  0.000 $\pm$  0.000 &  0.000 $\pm$  0.000\\ 
iSDSS &  0.007 $\pm$  0.010 &  0.015 $\pm$  0.022 &  0.093 $\pm$  0.021 &  0.014 $\pm$  0.031\\ 
\hline
\end{tabular}
\label{table:zp_offsets}
\end{table*}

 \begin{table*}[]
 \centering
 \caption{Stellar population properties of model templates\label{table:model-properties}}
 \begin{tabular}{|r|r|r|r|r|r|r|r|r|r|r|r|r|}
 \hline
  \multicolumn{1}{|c|}{\#} &
  \multicolumn{1}{c|}{$f_{dust}$} &
  \multicolumn{1}{c|}{$E$($B$-$V$)} &
  \multicolumn{1}{c|}{$\log U$} &
  \multicolumn{1}{c|}{$t_{burst}$} &
  \multicolumn{1}{c|}{$t_{main}$} &
  \multicolumn{1}{c|}{$f_{burst}$} &
  \multicolumn{1}{c|}{$\tau_{burst}$} &
  \multicolumn{1}{c|}{$\tau_{main}$} &
  \multicolumn{1}{c|}{$t_{mass}$} &
  \multicolumn{1}{c|}{$Z$} &
  \multicolumn{1}{c|}{SFR} &
  \multicolumn{1}{c|}{$M_*$} \\
  \multicolumn{1}{|c|}{} &
  \multicolumn{1}{c|}{} &
  \multicolumn{1}{c|}{[mag]} &
  \multicolumn{1}{c|}{} &
  \multicolumn{1}{c|}{[Myr]} &
  \multicolumn{1}{c|}{[Myr]} &
  \multicolumn{1}{c|}{} &
  \multicolumn{1}{c|}{[Myr]} &
  \multicolumn{1}{c|}{[Myr]} &
  \multicolumn{1}{c|}{[Myr]} &
  \multicolumn{1}{c|}{} &
  \multicolumn{1}{c|}{[M$_\odot$ yr$^{-1}$]} &
  \multicolumn{1}{c|}{[10$^9$ M$_\odot$]} \\
 \hline
01 & 0.00 & 0.2 & -3.0 &   100 &  2000 & 0.020 &    50 &  2000 &   729 & 0.008 & 2.67 & 2.28\\ 
02 & 0.50 & 0.5 & -3.0 &   200 &  5000 & 0.050 &    50 &  2000 &  2224 & 0.020 & 2.15 & 8.05\\ 
03 & 0.00 & 0.2 & -3.0 &   500 &  8000 & 0.050 &   200 &   500 &  6576 & 0.020 & 0.57 & 13.42\\ 
04 & 0.10 & 0.0 & -3.0 &   200 &  5000 & 0.020 &   100 &  2000 &  2297 & 0.008 & 0.72 & 2.38\\ 
05 & 0.20 & 0.5 & -3.0 &   100 &  8000 & 0.050 &    50 &  2000 &  4238 & 0.008 & 1.16 & 3.24\\ 
06 & 0.00 & 0.3 & -1.0 &   500 &  8000 & 0.020 &    50 &   500 &  6833 & 0.020 & 0.01 & 251.58\\ 
07 & 0.20 & 0.2 & -3.0 &  1000 &  8000 & 0.100 &   200 &  2000 &  4130 & 0.020 & 3.44 & 47.90\\ 
08 & 0.00 & 0.2 & -3.0 &   500 &  5000 & 0.050 &   200 &  1000 &  2961 & 0.020 & 0.92 & 9.33\\ 
09 & 0.20 & 0.5 & -3.0 &   100 &  5000 & 0.020 &    50 &  2000 &  2298 & 0.020 & 12.44 & 35.32\\ 
10 & 0.10 & 0.4 & -3.0 &   500 &  8000 & 0.010 &   200 &   500 &  6909 & 0.008 & 0.65 & 75.24\\ 
11 & 0.95 & 0.3 & -3.0 &  1000 &  8000 & 0.100 &    50 &   500 &  6300 & 0.020 & 0.00 & 18.86\\ 
12 & 0.00 & 0.0 & -3.0 &  1000 & 10000 & 0.020 &   100 &  1000 &  7818 & 0.020 & 0.02 & 18.46\\ 
13 & 0.95 & 0.0 & -3.0 &   100 &  8000 & 0.000 &    50 &   500 &  6995 & 0.008 & 0.00 & 32.42\\ 
14 & 0.00 & 0.3 & -3.0 &   200 &  8000 & 0.100 &   100 &  2000 &  3977 & 0.008 & 7.46 & 21.40\\ 
15 & 0.10 & 0.4 & -3.0 &   100 &  2000 & 0.050 &    50 &  1000 &   828 & 0.020 & 2.30 & 2.46\\ 
16 & 0.20 & 0.2 & -1.0 &   500 &  8000 & 0.010 &    50 &   500 &  6913 & 0.008 & 0.00 & 49.76\\ 
17 & 0.80 & 0.3 & -3.0 &   500 &  8000 & 0.005 &   100 &  1000 &  5958 & 0.020 & 1.19 & 208.87\\ 
18 & 0.10 & 0.0 & -3.0 &   500 & 10000 & 0.010 &   100 &  2000 &  6192 & 0.020 & 0.22 & 6.49\\ 
19 & 0.10 & 0.5 & -3.0 &   500 &  5000 & 0.050 &    50 &  2000 &  2251 & 0.020 & 3.28 & 13.89\\ 
20 & 0.10 & 0.5 & -3.0 &   200 &  8000 & 0.050 &    50 &  2000 &  4272 & 0.020 & 8.52 & 82.49\\ 
21 & 0.10 & 0.5 & -3.0 &   200 &  5000 & 0.050 &    50 &  1000 &  2941 & 0.020 & 4.39 & 47.56\\ 
22 & 0.50 & 0.3 & -3.0 &   100 & 10000 & 0.100 &   200 &  1000 &  6827 & 0.004 & 0.02 & 0.01\\ 
23 & 0.20 & 0.2 & -3.0 &   100 &  2000 & 0.020 &    50 &   500 &  1108 & 0.020 & 3.53 & 9.74\\ 
24 & 0.50 & 0.3 & -3.0 &   200 &  8000 & 0.005 &   200 &   500 &  6947 & 0.020 & 0.59 & 103.01\\ 
25 & 0.20 & 0.2 & -3.0 &   200 &  2000 & 0.100 &    50 &  1000 &   796 & 0.020 & 2.40 & 3.39\\ 
26 & 0.20 & 0.4 & -3.0 &   200 &  2000 & 0.050 &   200 &   500 &  1074 & 0.020 & 3.99 & 8.17\\ 
27 & 0.20 & 0.4 & -3.0 &   200 &  5000 & 0.050 &   100 &  2000 &  2220 & 0.020 & 7.19 & 19.49\\ 
28 & 0.20 & 0.0 & -3.0 &   200 &  5000 & 0.020 &    50 &   500 &  3892 & 0.020 & 0.01 & 0.65\\ 
29 & 0.00 & 0.4 & -1.0 &  1000 &  5000 & 0.100 &   100 &   500 &  3644 & 0.004 & 0.07 & 46.41\\ 
30 & 0.00 & 0.1 & -3.0 &   200 &  5000 & 0.020 &    50 &  1000 &  3052 & 0.020 & 0.90 & 12.07\\ 
31 & 0.00 & 0.3 & -3.0 &   500 &  5000 & 0.050 &   100 &  1000 &  2968 & 0.020 & 0.76 & 11.62\\ 
32 & 0.95 & 0.2 & -3.0 &   500 &  8000 & 0.005 &   200 &  1000 &  5957 & 0.020 & 0.57 & 61.60\\ 
33 & 0.10 & 0.5 & -3.0 &   200 &  2000 & 0.100 &   100 &  2000 &   677 & 0.008 & 2.16 & 1.77\\ 
34 & 0.50 & 0.5 & -3.0 &   200 &  2000 & 0.100 &   100 &   500 &  1018 & 0.020 & 7.77 & 15.89\\ 
35 & 0.00 & 0.1 & -2.5 &   500 &  8000 & 0.010 &   100 &  1000 &  5924 & 0.020 & 0.37 & 59.19\\ 
36 & 0.00 & 0.2 & -2.5 &  1000 &  8000 & 0.005 &   200 &   500 &  6958 & 0.008 & 0.05 & 147.79\\ 
37 & 0.20 & 0.0 & -3.0 &   500 &  5000 & 0.100 &   200 &  1000 &  2794 & 0.008 & 0.22 & 1.65\\ 
38 & 0.00 & 0.3 & -3.0 &   500 &  8000 & 0.005 &   100 &  1000 &  5958 & 0.020 & 1.11 & 194.79\\ 
39 & 0.20 & 0.3 & -3.0 &   200 & 10000 & 0.005 &    50 &  2000 &  6224 & 0.004 & 1.44 & 38.80\\ 
40 & 0.00 & 0.3 & -3.0 &  1000 &  8000 & 0.100 &   200 &  2000 &  4130 & 0.020 & 0.66 & 9.15\\ 
41 & 0.10 & 0.5 & -3.0 &   500 &  5000 & 0.020 &   200 &   500 &  3901 & 0.020 & 2.02 & 111.96\\ 
42 & 0.00 & 0.3 & -1.0 &  1000 &  5000 & 0.050 &    50 &   500 &  3820 & 0.004 & 0.01 & 3.65\\ 
43 & 0.95 & 0.0 & -3.0 &   100 & 10000 & 0.000 &    50 &   500 &  8996 & 0.008 & 0.00 & 22.92\\ 
44 & 0.00 & 0.2 & -2.5 &  1000 &  8000 & 0.050 &    50 &  1000 &  5704 & 0.004 & 0.04 & 8.11\\ 
45 & 0.00 & 0.3 & -3.0 &  1000 &  8000 & 0.100 &   100 &   500 &  6291 & 0.020 & 0.00 & 33.77\\ 
46 & 0.00 & 0.0 & -3.0 &  1000 & 10000 & 0.020 &    50 &  1000 &  7819 & 0.008 & 0.00 & 2.76\\ 
47 & 0.00 & 0.3 & -3.0 &   200 &  8000 & 0.005 &    50 &  1000 &  5953 & 0.008 & 0.07 & 7.95\\ 
48 & 0.00 & 0.5 & -3.0 &   200 &  5000 & 0.100 &   200 &  2000 &  2081 & 0.020 & 14.56 & 20.45\\ 
49 & 0.00 & 0.2 & -3.0 &  1000 &  8000 & 0.050 &   200 &  1000 &  5691 & 0.020 & 0.13 & 15.82\\ 
50 & 0.00 & 0.3 & -3.0 &   200 & 10000 & 0.020 &    50 &  1000 &  7771 & 0.008 & 1.06 & 68.77\\ 
\hline
\end{tabular}
\end{table*}

 \begin{table*}[]
 \centering
  \caption{Observables measured on model templates\label{table:observables}}
 \begin{tabular}{|r|r|r|r|r|r|r|r|r|}
 \hline
  \multicolumn{1}{|c|}{\#} &
  \multicolumn{1}{c|}{$\log L_\nu$($g$)} &
  \multicolumn{1}{c|}{$u$-$r$} &
  \multicolumn{1}{c|}{$g$-$i$} &
  \multicolumn{1}{c|}{$\beta$} &
  \multicolumn{1}{c|}{$D_n$(4000)} &
  \multicolumn{1}{c|}{EW([O {\sc ii}])} &
  \multicolumn{1}{c|}{EW([O {\sc iii}])} &
  \multicolumn{1}{c|}{EW(H$\alpha$)} \\
  \multicolumn{1}{|c|}{} &
  \multicolumn{1}{c|}{[W m$^{-2}$]} &
  \multicolumn{1}{c|}{[AB mag]} &
  \multicolumn{1}{c|}{[AB mag]} &
  \multicolumn{1}{c|}{} &
  \multicolumn{1}{c|}{} &
  \multicolumn{1}{c|}{[nm]} &
  \multicolumn{1}{c|}{[nm]} &
  \multicolumn{1}{c|}{[nm]} \\
 \hline
01 & 21.69 & 0.82 & 0.23 & -2.04 & 1.12 & 6.915 & 2.788 & 11.634\\ 
02 & 21.51 & 1.59 & 0.83 & -1.44 & 1.25 & 0.775 & 0.208 & 1.488\\ 
03 & 21.46 & 1.67 & 0.83 & -1.73 & 1.32 & 2.511 & 0.439 & 2.262\\ 
04 & 21.46 & 0.90 & 0.32 & -2.24 & 1.17 & 6.636 & 2.024 & 6.495\\ 
05 & 21.14 & 1.30 & 0.68 & -1.66 & 1.18 & 2.363 & 1.118 & 4.762\\ 
06 & 22.41 & 2.59 & 1.24 &  0.85 & 1.70 & 0.008 & 0.052 & 0.072\\ 
07 & 22.19 & 1.70 & 0.83 & -1.80 & 1.36 & 1.968 & 0.346 & 1.790\\ 
08 & 21.58 & 1.56 & 0.77 & -1.80 & 1.30 & 2.899 & 0.545 & 2.924\\ 
09 & 22.15 & 1.49 & 0.84 & -1.53 & 1.23 & 1.780 & 0.520 & 3.638\\ 
10 & 21.90 & 2.39 & 1.26 & -1.33 & 1.59 & 0.861 & 0.188 & 0.496\\ 
11 & 21.38 & 2.53 & 1.19 &  3.50 & 1.72 & 0.004 & 0.000 & 0.002\\ 
12 & 21.44 & 2.35 & 1.09 & -0.85 & 1.82 & 0.618 & 0.057 & 0.199\\ 
13 & 21.77 & 2.22 & 1.06 &  0.84 & 1.78 & 0.015 & 0.002 & 0.005\\ 
14 & 22.19 & 1.08 & 0.46 & -1.86 & 1.16 & 4.293 & 1.739 & 6.975\\ 
15 & 21.43 & 1.15 & 0.55 & -1.67 & 1.17 & 2.766 & 0.858 & 6.533\\ 
16 & 21.85 & 2.35 & 1.15 &  0.07 & 1.68 & 0.007 & 0.092 & 0.080\\ 
17 & 22.35 & 2.53 & 1.28 & -1.37 & 1.72 & 0.105 & 0.013 & 0.053\\ 
18 & 21.24 & 1.63 & 0.81 & -2.09 & 1.39 & 3.220 & 0.458 & 1.918\\ 
19 & 21.69 & 1.71 & 0.89 & -1.47 & 1.28 & 1.868 & 0.467 & 3.175\\ 
20 & 22.30 & 1.73 & 0.91 & -1.34 & 1.26 & 1.208 & 0.299 & 2.015\\ 
21 & 22.09 & 1.82 & 0.96 & -1.31 & 1.29 & 1.060 & 0.248 & 1.618\\ 
22 & 19.11 & 0.69 & 0.26 & -2.16 & 1.08 & 2.322 & 1.896 & 8.827\\ 
23 & 21.97 & 1.22 & 0.55 & -1.89 & 1.22 & 2.628 & 0.609 & 3.789\\ 
24 & 22.05 & 2.43 & 1.24 & -1.35 & 1.70 & 0.115 & 0.012 & 0.055\\ 
25 & 21.71 & 1.00 & 0.38 & -1.89 & 1.17 & 2.919 & 0.758 & 5.317\\ 
26 & 21.77 & 1.38 & 0.69 & -1.64 & 1.22 & 2.058 & 0.555 & 3.899\\ 
27 & 22.02 & 1.35 & 0.70 & -1.65 & 1.21 & 2.070 & 0.573 & 3.957\\ 
28 & 20.34 & 1.69 & 0.80 & -1.86 & 1.39 & 1.063 & 0.150 & 0.638\\ 
29 & 22.02 & 2.26 & 1.08 &  1.14 & 1.50 & 0.003 & 0.080 & 0.095\\ 
30 & 21.73 & 1.47 & 0.72 & -1.91 & 1.30 & 2.945 & 0.522 & 2.596\\ 
31 & 21.55 & 1.84 & 0.91 & -1.56 & 1.36 & 1.889 & 0.338 & 1.850\\ 
32 & 21.91 & 2.34 & 1.18 & -1.70 & 1.66 & 0.042 & 0.005 & 0.021\\ 
33 & 21.35 & 1.22 & 0.50 & -1.63 & 1.17 & 3.207 & 1.501 & 7.322\\ 
34 & 22.02 & 1.47 & 0.74 & -1.46 & 1.23 & 0.822 & 0.236 & 1.794\\ 
35 & 22.00 & 2.22 & 1.06 & -1.57 & 1.62 & 0.610 & 0.227 & 0.500\\ 
36 & 22.29 & 2.41 & 1.20 &  0.25 & 1.77 & 0.177 & 0.111 & 0.138\\ 
37 & 21.17 & 1.15 & 0.42 & -2.11 & 1.24 & 3.824 & 1.015 & 3.133\\ 
38 & 22.32 & 2.51 & 1.27 & -1.35 & 1.71 & 0.773 & 0.094 & 0.391\\ 
39 & 21.95 & 1.89 & 0.95 & -1.67 & 1.39 & 1.052 & 0.407 & 1.279\\ 
40 & 21.39 & 1.81 & 0.91 & -1.66 & 1.37 & 2.274 & 0.419 & 2.271\\ 
41 & 22.11 & 2.47 & 1.32 & -1.21 & 1.55 & 0.698 & 0.111 & 0.560\\ 
42 & 20.94 & 2.20 & 1.05 &  1.14 & 1.54 & 0.005 & 0.105 & 0.113\\ 
43 & 21.53 & 2.29 & 1.10 &  0.39 & 1.84 & 0.017 & 0.002 & 0.005\\ 
44 & 21.28 & 2.07 & 0.98 & -0.45 & 1.51 & 0.185 & 0.221 & 0.346\\ 
45 & 21.65 & 2.50 & 1.17 &  2.67 & 1.69 & 0.113 & 0.014 & 0.063\\ 
46 & 20.73 & 2.13 & 1.00 & -0.25 & 1.70 & 0.627 & 0.098 & 0.207\\ 
47 & 21.08 & 2.20 & 1.14 & -1.43 & 1.54 & 0.998 & 0.217 & 0.561\\ 
48 & 22.10 & 1.25 & 0.68 & -1.56 & 1.16 & 2.919 & 0.983 & 7.567\\ 
49 & 21.41 & 2.26 & 1.09 & -1.34 & 1.61 & 0.931 & 0.117 & 0.503\\ 
50 & 22.07 & 1.84 & 0.95 & -1.56 & 1.35 & 1.223 & 0.330 & 0.974\\ 

\hline
\end{tabular}
\end{table*}


\begin{thebibliography}{999}
\bibitem[Alarcon et al.(2021)]{Alarcon21} Alarcon, A., Gaztanaga, E., Eriksen, M., et al.\ 2021, \mnras, 501, 6103
\bibitem[Angulo et al.(2008)]{Angulo08} Angulo, R.~E., Baugh, C.~M., Frenk, C.~S., et al.\ 2008, MNRAS, 383, 755
\bibitem[Anvi(1976)]{Anvi76}Avni Y., 1976, ApJ, 210, 642
\bibitem[Arnouts, \& Ilbert(2011)]{Arnouts11}Arnouts, S., \& Ilbert, O.\ 2011, LePHARE: Photometric Analysis for Redshift Estimate, ascl:1108.009
\bibitem[Balogh et al.(1999)]{Balogh99} Balogh, M.~L., Morris, S.~L., Yee, H.~K.~C., et al.\ 1999, ApJ, 527, 54
\bibitem[Baqui et al.(2020)]{Baqui20}Baqui, P.~O., Marra, V., Casarini, L. et al. 2020, A\&A submitted, arxiv:2007.07622
\bibitem[Baum(1962)]{Baum62}Baum, W.~A.\ 1962, Problems of Extra-Galactic Research, 15, 390
\bibitem[Ben{\'i}tez et al.(2009)]{Benitez09}Ben\'itez, N., Moles, M., Aguerri, J. A. L., et al. 2009, ApJ, 692, L5
\bibitem[Ben{\'i}tez et al.(2014)]{Benitez14}Ben\'itez, N., Dupke, R., Moles, M., et al. 2014, arXiv e-prints, arXiv:1403.5237
\bibitem[Ben{\'i}tez(2000)]{Benitez00}Ben{\'\i}tez, N.\ 2000, ApJ, 536, 571
\bibitem[Bertin \& Arnouts(1996)]{Bertin96}Bertin, E. \& Arnouts, S. 1996, A\&AS, 117, 393
\bibitem[Bertin et al.(2002)]{Bertin02} Bertin, E., Mellier, Y., Radovich, M., et al.\ 2002, Astronomical Data Analysis Software and Systems XI, 281, 228
\bibitem[Bertin(2006)]{Bertin06} Bertin, E.\ 2006, Astronomical Data Analysis Software and Systems XV, 351, 112
\bibitem[Blake \& Bridle(2005)]{Blake05} Blake, C. \& Bridle, S.\ 2005, MNRAS, 363, 1329
\bibitem[Bolzonella et al.(2000)]{Bolzonella00}Bolzonella M., Miralles J. M., Pell\'o R., 2000, A\&A, 363, 476
\bibitem[Bonoli et al.(2020)]{Bonoli21} Bonoli, S., Mar{\'\i}n-Franch, A., Varela, J., et al.\ 2020, A\&A, in press. arXiv:2007.01910
\bibitem[Boquien et al.(2019)]{Boquien19} Boquien, M., Burgarella, D., Roehlly, Y., et al.\ 2019, A\&A, 622, A103
\bibitem[Brammer et al.(2008)]{Brammer08} Brammer, G.~B., van Dokkum, P.~G., \& Coppi, P.\ 2008, ApJ, 686, 1503
\bibitem[Bruzual \& Charlot(2003)]{Bruzual03} Bruzual, G., \& Charlot, S.\ 2003, MNRAS, 344, 1000
\bibitem[Calzetti et al.(2000)]{Calzetti00} Calzetti, D., Armus, L., Bohlin, R.~C., et al.\ 2000, ApJ, 533, 682
\bibitem[Chabrier(2003)]{Chabrier03} Chabrier, G.\ 2003, PASP, 115, 763
\bibitem[Chaves-Montero et al.(2018)]{Chaves-Montero18} Chaves-Montero, J., Angulo, R.~E., \& Hern{\'a}ndez-Monteagudo, C.\ 2018, MNRAS, 477, 3892
\bibitem[Coe et al.(2006)]{Coe06} Coe, D., Ben{\'\i}tez, N., S{\'a}nchez, S.~F., et al.\ 2006, AJ, 132, 926
\bibitem[Coil et al.(2004)]{Coil04} Coil, A.~L., Newman, J.~A., Kaiser, N., et al.\ 2004, ApJ, 617, 765
\bibitem[Cooper et al.(2011)]{Cooper11} Cooper, M.~C., Aird, J.~A., Coil, A.~L., et al.\ 2011, ApJS, 193, 14
\bibitem[Couch et al.(1983)]{Couch83} Couch, W.~J., Ellis, R.~S., Godwin, J., et al.\ 1983, MNRAS, 205, 1287
\bibitem[Dahlen et at.(2013)]{Dahlen13}Dahlen T. et al., 2013, ApJ, 775, 93
\bibitem[Davis et al.(2003)]{Davis03} Davis, M., Faber, S.~M., Newman, J., et al.\ 2003, SPIE, 4834, 161
\bibitem[Eriksen et al.(2019)]{Eriksen19}Eriksen, M., Alarcon, A., Gaztanaga, E., et al. 2019, MNRAS, 484, 4200
\bibitem[Ferland et al.(1998)]{Ferland98} Ferland, G.~J., Korista, K.~T., Verner, D.~A., et al.\ 1998, PASP, 110, 761
\bibitem[Ferland et al.(2013)]{Ferland13} Ferland, G.~J., Porter, R.~L., van Hoof, P.~A.~M., et al.\ 2013, RMxAA, 49, 137
\bibitem[Fern{\'a}ndez-Soto et al.(1999)]{Fernandez-Soto99} Fern{\'a}ndez-Soto, A., Lanzetta, K.~M., \& Yahil, A.\ 1999, ApJ, 513, 34
\bibitem[Fern{\'a}ndez-Soto et al.(2002)]{Fernandez-Soto02}Fern\'andez-Soto A., Lanzetta K. M., Chen H.-W., Levine B., Yahata N., 2002, MNRAS, 330, 889
\bibitem[Gonz{\'a}lez Delgado et al.(2021)]{GonzalezDelgado21} Gonz{\'a}lez Delgado, R.~M., D{\'\i}az-Garc{\'\i}a, L.~A., de Amorim, A., et al.\ 2021, A\&A, in press. arXiv:2102.13121
\bibitem[Green et al.(2018)]{Green18}Green, G. M., Schlafly, E. F., Finkbeiner, D., et al. 2018, MNRAS, 478, 651
\bibitem[Hern\'an-Caballero et al.(2013)]{Hernan-Caballero13}Hern\'an-Caballero A., et al., 2013, MNRAS, 434, 2136
\bibitem[Hern{\'a}n-Caballero et al.(2015)]{Hernan-Caballero15} Hern{\'a}n-Caballero, A., Alonso-Herrero, A., Hatziminaoglou, E., et al.\ 2015, ApJ, 803, 109
\bibitem[Hern{\'a}n-Caballero(2012)]{Hernan-Caballero12}Hern\'an-Caballero, A. 2012, MNRAS, 427, 816
\bibitem[Hildebrandt et al.(2008)]{Hildebrandt08} Hildebrandt, H., Wolf, C., \& Ben{\'\i}tez, N.\ 2008, A\&A, 480, 703
\bibitem[Ilbert et al.(2006)]{Ilbert06}Ilbert, O., Arnouts, S., McCracken, H.~J., et al.\ 2006, A\&A, 457, 841
\bibitem[Inoue(2011)]{Inoue11} Inoue, A.~K.\ 2011, MNRAS, 415, 2920
\bibitem[Kauffmann et al.(2003a)]{Kauffmann03a}Kauffmann G., et al. 2003a, MNRAS, 341, 33
\bibitem[Kauffmann et al.(2003b)]{Kauffmann03b}Kauffmann G., et al. 2003b, MNRAS, 341, 54
\bibitem[Kriek et al.(2006)]{Kriek06}Kriek M., et al., 2006, ApJ, 645, 44
\bibitem[Kron(1980)]{Kron80}Kron, R. G., 1980, ApJS, 43, 305
\bibitem[Lanzetta et al.(1996)]{Lanzetta96} Lanzetta, K.~M., Yahil, A., \& Fern{\'a}ndez-Soto, A.\ 1996, Nature, 381, 759
\bibitem[L{\'o}pez-Sanjuan et al.(2019a)]{Lopez-Sanjuan19a} L{\'o}pez-Sanjuan, C., V{\'a}zquez Rami{\'o}, H., Varela, J., et al.\ 2019a, A\&A, 622, A177 
\bibitem[L{\'o}pez-Sanjuan et al.(2019b)]{Lopez-Sanjuan19b}L\'opez-Sanjuan, C., Varela, J., Crist\'obal-Hornillos, D., et al. 2019b, A\&A, 631, A119
\bibitem[Le F{\`e}vre et al.(2005)]{LeFevre05} Le F{\`e}vre, O., Vettolani, G., Garilli, B., et al.\ 2005, \aap, 439, 845
\bibitem[Mobasher et al.(1996)]{Mobasher96} Mobasher, B., Rowan-Robinson, M., Georgakakis, A., et al.\ 1996, \mnras, 282, L7
\bibitem[Moles et al.(2008)]{Moles08}Moles, M., Ben\'itez, N., Aguerri, J. A. L., et al. 2008, AJ, 136, 1325
\bibitem[Molino et al.(2014)]{Molino14}Molino, A., Ben\'itez, N., Moles, M., et al. 2014, MNRAS, 441, 2891
\bibitem[Molino et al.(2019)]{Molino19}Molino, A., Costa-Duarte, M. V., Mendes de Oliveira, C., et al. 2019, A\&A, 622, A178
\bibitem[Molino et al.(2017)]{Molino17}Molino, A., Ben\'itez, N., Ascaso, B., et al. 2017, MNRAS, 470, 95
\bibitem[Mundy et al.(2017)]{Mundy17} Mundy, C.~J., Conselice, C.~J., Duncan, K.~J., et al.\ 2017, MNRAS, 470, 3507
\bibitem[Newman et al.(2013)]{Newman13} Newman, J.~A., Cooper, M.~C., Davis, M., et al.\ 2013, ApJS, 208, 5
\bibitem[Noll et al.(2009)]{Noll09} Noll, S., Burgarella, D., Giovannoli, E., et al.\ 2009, A\&A, 507, 1793
\bibitem[Oyaizu(2008)]{Oyaizu08}Oyaizu H., Lima M., Cunha C. E., Lin H., Frieman J., 2008, ApJ, 689, 709
\bibitem[P{\'e}rez-Gonz{\'a}lez et al.(2013)]{Perez-Gonzalez13}P\'erez-Gonz\'alez, P. G., Cava, A., Barro, G., et al. 2013, ApJ, 762, 46
\bibitem[Polsterer et al.(2016)]{Polsterer16} Polsterer, K.~L., D'Isanto, A., \& Gieseke, F.\ 2016, arXiv:1608.08016
\bibitem[Press et al.(1992)]{Press92}Press W. H., Teukolsky S. A., Vetterling W. T., Flannery B. P., 1992, Numerical Recipes in C: The Art of Scientific Computing, 2nd edn. Cambridge Univ. Press, Cambridge
\bibitem[Salvato et al.(2019)]{Salvato19} Salvato, M., Ilbert, O., \& Hoyle, B.\ 2019, Nature Astronomy, 3, 212
\bibitem[Schlafly et al.(2016)]{Schlafly16}Schlafly, E. F., Meisner, A. M., Stutz, A. M., et al. 2016, ApJ, 821, 78
\bibitem[Schmidt et al.(2020)]{Schmidt20} Schmidt, S.~J., Malz, A.~I., Soo, J.~Y.~H., et al.\ 2020, MNRAS, 499, 1587
\bibitem[Schmidt \& Thorman(2013)]{Schmidt13}Schmidt S. J., Thorman P., 2013, MNRAS, 431, 2766
\bibitem[Taniguchi et al.(2007)]{Taniguchi07} Taniguchi, Y., Scoville, N., Murayama, T., et al.\ 2007, ApJS, 172, 9
\bibitem[Taniguchi et al.(2015)]{Taniguchi15} Taniguchi, Y., Kajisawa, M., Kobayashi, M.~A.~R., et al.\ 2015, PASJ, 67, 104
\bibitem[Whitten et al.(2019)]{Whitten19}Whitten, D. D., Placco, V. M., Beers, T. C., et al. 2019, A\&A, 622, A182
\bibitem[Wittman et al.(2016)]{Wittman16}Wittman D., Bhaskar R., Tobin R., 2016, MNRAS, 457, 4005
\bibitem[Wolf et al.(2003)]{Wolf03}Wolf, C., Meisenheimer, K., Rix, H. W., et al. 2003, A\&A, 401, 73
\bibitem[York et al.(2000)]{York00}York, D. G., Adelman, J., Anderson, John E., J., et al. 2000, AJ, 120, 1579

\end{thebibliography}
\end{document}